\documentclass[journal]{IEEEtran}

\usepackage{amsmath}
\usepackage{amssymb}
\usepackage{array}
\usepackage{booktabs}
\usepackage{cite}
\usepackage{graphicx}
\usepackage{tabularx}
\usepackage{url}


\newcolumntype{Y}{>{\raggedright\arraybackslash}X}
\newcolumntype{L}[1]{>{\raggedright\arraybackslash}p{#1}}

\begin{document}

\title{Single-Photon Detectors for Optical Wireless Communications: A Tutorial}
\author{Ziyuan~Shi,~\IEEEmembership{Senior Member,~IEEE},
Yueying~Zhan,
Lei~Yang,
and Mohamed-Slim~Alouini,~\IEEEmembership{Fellow,~IEEE}%
\thanks{(\textit{*Corresponding author:Yueying Zhan*})}%
\thanks{Z. Shi and M.-S. Alouini are with the Computer, Electrical,
and Mathematical Science and Engineering (CEMSE) Division, King Abdullah
University of Science and Technology (KAUST), Thuwal, Makkah Province,
Saudi Arabia (e-mail: ziyuan.shi@kaust.edu.sa;
slim.alouini@kaust.edu.sa).}%
\thanks{Y. Zhan and L. Yang are with the Technology and Engineering
Center for Space Utilization, Chinese Academy of Sciences (CSU),
Beijing 100094, China (e-mail: zhanyueying@csu.ac.cn;
yang.lei@csu.ac.cn).}%
}
\maketitle

\begin{abstract}
Photon-counting receivers extend optical wireless communication (OWC) into
regimes in which only a few detected photons are available per information
bit, but the detector can no longer be treated as an interchangeable ideal
counter.  This tutorial develops a communication-oriented, port-to-bit
framework for single-photon detection in indoor, terrestrial, airborne,
space, underwater, non-line-of-sight, and integrated communication-and-sensing
links.  It compares SPADs, silicon photomultipliers, CMOS SPAD arrays,
superconducting nanowire single-photon detectors, PMT/MCP-PMT receivers,
upconversion detectors, and emerging technologies through photon-detection
efficiency, timing, active area, dynamic range, spectral coverage, readout,
cooling, and integration requirements.  Beginning with the ideal Poisson
channel, the tutorial then develops communication-level models for dead time,
recovery, finite-pixel occupancy, saturation, background counts, afterpulsing,
crosstalk, nonuniformity, and event loss.  These models are connected to
modulation, likelihood-based detection, synchronization, channel estimation,
soft decoding, and detector-aware receiver processing, with numerical studies
that compare algorithms under a common detector and detectors under a common
algorithm.  Finally, a unified link-budget and system co-design methodology is
used to relate photons per bit, sensitivity, bandwidth, data rate, and link
margin, and representative demonstrations are normalized to expose practical
tradeoffs.  The central conclusion is that useful photon efficiency is an
end-to-end property of the optical front end, detector, readout, and digital
receiver, rather than a detector metric in isolation.
\end{abstract}

\begin{IEEEkeywords}
Channel estimation, CMOS SPAD array, free-space optical communication,
optical wireless communication, photon counting, silicon photomultiplier
(SiPM), single-photon avalanche diode (SPAD), superconducting nanowire
single-photon detector (SNSPD), synchronization.
\end{IEEEkeywords}

\section{Introduction}
\label{sec:introduction}

\IEEEPARstart{O}{ptical} wireless communication (OWC) has developed into a
family of complementary technologies rather than a single replacement for
radio-frequency links.  Indoor visible-light and near-infrared systems exploit
unlicensed optical bandwidth for access, positioning, and device connectivity;
terrestrial free-space optical (FSO) links provide high-capacity point-to-point
 backhaul \cite{Haas2020OWC,Khalighi2014FSOSurvey,Mapunda2020IndoorVLC,He2023VLC}.
 Airborne, satellite, and deep-space terminals use narrow optical beams to
 overcome extreme range, while underwater or non-line-of-sight (NLOS) systems
 operate where radio or conventional line-of-sight optics are ineffective
 \cite{Wu2017DeepSpace,Hu2018UWOC256PPM,Chen2020LongDistanceUWOC,Gong2015NLOS}.
Although these links differ greatly in wavelength, range, propagation loss,
background environment, and platform constraints, they increasingly encounter
the same receiver question: how much reliable information can be recovered
when only a small and random number of photons reaches the detector during each
useful observation interval?

Photon counting changes both the sensitivity limit and the mathematical form
of this question.  Instead of observing a nearly continuous photocurrent with
approximately additive Gaussian noise, the receiver observes discrete events
whose number, arrival time, and possibly spatial coordinates carry information.
The ideal reference is a Poisson channel, but the practical objective is not
merely to register an isolated photon; it is to transform a photon-starved
optical waveform into synchronized, decoded bits at a useful rate and with a
declared error probability.  Classical photon-counting analyses established
the statistical foundations of this regime \cite{Karp1970PhotonCounting,Yiannopoulos2019ErrorProbability}, while photon-efficient modulation and coding
show how high information efficiency can be approached with sparse optical
signals \cite{Caplan2006PhotonEfficient,Dolinar2012UltimateLimits,Banaszek2020QuantumLimits}.  Operational and field-oriented demonstrations,
including the Lunar Laser Communication Demonstration and current deep-space
 ground-receiver programs, confirm that single-photon reception is a system
 technology rather than only a laboratory detector result
 \cite{Grein2015LLCDReceiver,Grein2015Moon,Srinivasan2023DSOCReceiver}.
 Recent deep-space and ground-receiver studies reinforce the same conclusion
 \cite{Biswas2024DeepSpacePhoton,Marsden2024GroundReceiver}.

\begin{figure*}[!t]
  \centering
  \includegraphics[width=0.98\textwidth]{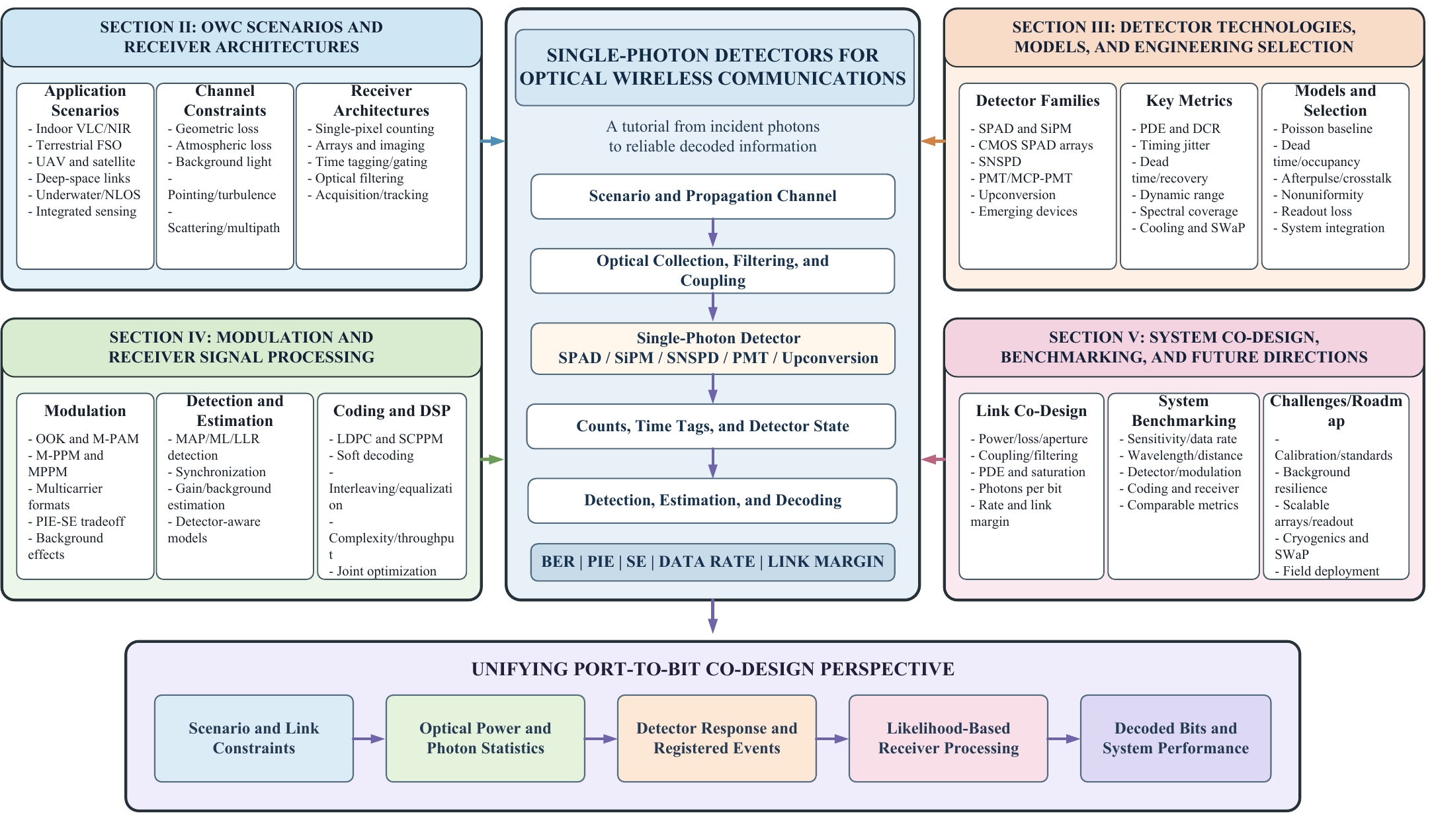}
  \caption{Organization and unifying port-to-bit co-design perspective of this
  tutorial.  Application and channel constraints determine the optical front
  end and detector operating point; the resulting counts, time tags, and
  detector state determine the likelihood used by synchronization, estimation,
  detection, and decoding.  The common comparison endpoint is reliable system
  performance rather than an isolated detector metric.}
  \label{fig:paper_overview}
\end{figure*}

Several detector families can provide this interface, but they expose very
different information and implementation constraints.  Single-photon
avalanche diodes (SPADs) and silicon photomultipliers (SiPMs) offer compact,
solid-state reception and can be integrated with quenching, time-to-digital
 conversion, and digital processing.  CMOS SPAD arrays add spatial resolution
 and massive parallelism \cite{Fisher2017CMOSSPAD,Li2022IntegratedSPAD,Liu2025NovelCMOSSPAD}.  Communication-oriented SPAD studies now span
 higher-order modulation, ACO-OFDM, information-rate analysis, and receiver
 operation without explicit photon counting
 \cite{Lit0018,Det0051,Lit0096,Lit0104}.  Superconducting nanowire single-photon detectors
(SNSPDs) combine low noise, high efficiency, and precise timing, and
multi-element arrays extend their count-rate and tracking capability, at the
cost of cryogenic infrastructure \cite{Zhang2021SNSPDMechanism,Hao2024SNSPDArray,Det0129,Wan2024DaytimeSNSPD}.  Photomultiplier tubes and microchannel-plate PMTs
remain attractive where active area, ultraviolet response, or fast timing is
more important than solid-state integration; upconversion receivers translate
telecommunication-band photons to wavelengths accessible to efficient silicon
detectors \cite{Farr2005Detectors,Farr2009NAF,Det0271}.  Emerging InGaAs SPADs,
negative-feedback avalanche devices, and heterogeneous or hybrid receivers
further blur the boundary between a detector and a complete front end
\cite{Lai2025InGaAsSPAD,Det0062}.

Device data sheets alone are therefore insufficient for receiver selection.
Photon-detection efficiency (PDE), dark-count rate, timing jitter, active area,
spectral coverage, dead time, and maximum count rate must be interpreted at the
operating bias, temperature, optical spot size, and electronic readout load.
Dead time and recharge make the count process state dependent; finite-pixel
occupancy compresses large signals; afterpulsing and crosstalk create correlated
events; pixel-to-pixel nonuniformity couples pointing motion to apparent gain;
and shared time-to-digital converters, arbiters, and interfaces can discard
events that were successfully detected.  Statistical SPAD and SiPM models have
quantified parts of this behavior \cite{Sarbazi2018SPADStatistics,Fisher2013IntegratingReceiver,Det0385,Det0390}, and communication experiments
have demonstrated the associated nonlinear distortion, error floors, and need
for operating-point control \cite{Huang2023SPADOFDM,Liu2023TimeControlledSPAD,Zhang2022SiPMGuideline,Det0123}.  Nevertheless, a detector with the largest
 isolated-device PDE can yield a poorer link if its dynamic range, recovery,
 coupling, or readout cannot support the selected waveform and symbol rate.
 Time-interleaved reception, mixed Poisson--Gaussian modeling, and array-level
 dead-time mitigation further show that the relevant state extends beyond a
 single junction \cite{Lit0118,Lit0161,Lit0381}.

The same dependence appears in receiver signal processing.  OOK, multilevel
PAM, PPM, and multicarrier formats distribute optical energy differently in
time and amplitude, so they interact differently with background counts,
saturation, and timing uncertainty.  Maximum-likelihood and multiple-symbol
detectors can exploit the correct photon-counting statistics
\cite{Chatzidiamantis2010MLSD,Riediger2007MSD}, but their likelihoods become
mismatched when a practical detector is replaced by an ideal Poisson counter.
Synchronization and channel estimation likewise depend on the detector's
timing response, count compression, and admitted background
\cite{Det0014,Det0156,Det0138}.  At the coding stage, soft information must be
calibrated to the actual count law, while decoding complexity and interface
 throughput constrain the realizable information rate
 \cite{Lit0358,Lit0367,Lit0371,Lit0373}.
These couplings motivate a communication-oriented treatment in which detector
physics and digital algorithms are evaluated against the same end-to-end
metrics.

Existing OWC surveys provide broad coverage of channels, networking, and
modulation \cite{Haas2020OWC,Khalighi2014FSOSurvey,Huang2023SixG}, whereas
detector-focused studies generally emphasize device physics or individual
figures of merit.  A complementary tutorial is needed to connect application
conditions, optical collection, detector response, event statistics,
inference, coding, and system performance without losing the nonidealities at
their interfaces.  This article adopts that \emph{port-to-bit} perspective:
optical power and background are defined at a declared receiver port, mapped
through the optical front end and detector to reported counts or time tags, and
then propagated through synchronization, estimation, detection, and decoding
to bit-error rate (BER), photon information efficiency (PIE), spectral
efficiency (SE), throughput, and link margin.  Figure~\ref{fig:paper_overview}
 summarizes this organizing principle and the relationship among the four main
 sections.
 Complementary surveys of hybrid OWC networks, optical IoT, advanced VLC
 transceivers, and laser-based VLC provide the broader system context in which
 the port-to-bit interface must operate \cite{Lit0198,Lit0203,Lit0212,Lit0245}.

The main contributions of this tutorial are fivefold.
\begin{itemize}
  \item It maps representative indoor, terrestrial, airborne, space,
  underwater, NLOS, and integrated sensing scenarios to photon-counting
  receiver architectures, making explicit how propagation, background, field
  of view, pointing, and SWaP constraints alter the preferred architecture.

  \item It compares SPADs, SiPMs, CMOS SPAD arrays, SNSPDs, PMT/MCP-PMTs,
  upconversion receivers, and emerging devices through communication-relevant
  metrics and through the complete path from an incident photon to a registered
  digital event.

  \item It develops a unified progression from the ideal Poisson channel to
  detector-aware models of dead time, recovery, occupancy, saturation,
  background, afterpulsing, crosstalk, nonuniformity, and readout loss, and then
  translates those effects into engineering selection and integration rules.

  \item It treats modulation, detection, synchronization, channel estimation,
  coding, and advanced receiver processing under consistent photon-counting
  statistics.  Comparative numerical studies expose both same-detector/
  different-algorithm and same-algorithm/different-detector tradeoffs rather
  than presenting isolated performance curves.

  \item It closes the analysis with a system-level link-budget and co-design
  framework, a normalized comparison of representative demonstrations, and a
  roadmap covering calibration, reporting standards, scalable readout,
  background resilience, SWaP, and field deployment.
\end{itemize}

The remainder of the article follows the structure in
Fig.~\ref{fig:paper_overview}.  Section~\ref{sec:scenarios_architectures}
reviews OWC scenarios and receiver architectures.  Section~\ref{sec:detector_technologies}
compares detector technologies, performance metrics, communication-oriented
nonideal models, and engineering selection.  Section~\ref{sec:modulation_signal_processing}
develops photon-counting modulation and receiver signal processing, including
detection, synchronization, estimation, and coding.  Finally,
Section~\ref{sec:system_level_codesign} addresses link budgeting, system-level
co-design, experimental benchmarking, open challenges, and research
directions.  Section~\ref{sec:conclusion} concludes the tutorial.

\section{Optical Wireless Communication Scenarios and Receiver Architectures}
\label{sec:scenarios_architectures}

Optical wireless communication (OWC) spans links that differ by more than ten
orders of magnitude in range, from room-scale visible-light links to deep-space
telemetry.  Consequently, ``a photon-counting receiver'' is not a single
architecture.  It is a chain of optical collection, spectral and spatial
filtering, photon detection, time tagging, synchronization, inference, and
decoding whose optimum partition depends on the propagation medium and the
platform.  General OWC and free-space optical (FSO) surveys provide the wider
communications context \cite{Haas2020OWC,Khalighi2014FSOSurvey}, while recent
work places SPAD and SNSPD receivers among the candidate technologies for
photon-starved future networks \cite{Huang2023SixG}.  The unifying benefit is
high sensitivity; the accompanying costs are Poisson statistics, background
counts, dead time, saturation, afterpulsing, timing jitter, and---for many
SNSPDs---cryogenic operation.  These constraints must therefore be assigned to
the complete receiver rather than to the detector alone.

For a direct-detection link, a useful first-order interface between propagation
and receiver design is the mean detected signal-photon number per observation
interval,
\begin{equation}
  \bar n_{\mathrm{s}}=
  \frac{P_{\mathrm{t}}T_{\mathrm{s}}}{h\nu}
  \eta_{\mathrm{tx}}\eta_{\mathrm{ch}}\eta_{\mathrm{col}}
  \eta_{\mathrm{filt}}\eta_{\mathrm{cpl}}\eta_{\mathrm{det}},
  \label{eq:mean_photons}
\end{equation}
where $P_{\mathrm{t}}$ and $T_{\mathrm{s}}$ are transmitted optical power and
symbol duration, $\eta_{\mathrm{ch}}$ includes geometric and medium losses,
$\eta_{\mathrm{col}}$ is the aperture collection efficiency, and the remaining
terms describe spectral filtering, spatial coupling, and photon-detection
efficiency.  In the unsaturated regime the observed count is commonly modeled
as $N\sim\mathrm{Poisson}(\bar n_{\mathrm{s}}+\bar n_{\mathrm{b}}+\bar
n_{\mathrm{d}})$, with background and dark-count means $\bar n_{\mathrm{b}}$
and $\bar n_{\mathrm{d}}$ \cite{Karp1970PhotonCounting,Yiannopoulos2019ErrorProbability}.
Equation~\eqref{eq:mean_photons} is deliberately modular: the scenarios below
mainly change $\eta_{\mathrm{ch}}$ and $\bar n_{\mathrm{b}}$, whereas receiver
architectures trade $\eta_{\mathrm{col}}\eta_{\mathrm{cpl}}$ against spatial
rejection, tracking observability, speed, and implementation burden.

\subsection{Representative Application Scenarios}

\subsubsection{Indoor visible-light and near-infrared communication}
Indoor OWC uses luminaires or dedicated emitters for access, localization, and
device connectivity.  Visible-light communication (VLC) benefits from existing
illumination infrastructure, whereas near-infrared (NIR) removes the illumination
constraint and can use narrower beams.  Reflections create a deterministic but
geometry-dependent multipath channel; sunlight, lamps, and display sources may
produce a background count rate much larger than the signal rate.  Broad-area
SPADs or SiPMs ease alignment and allow compact non-imaging receivers, while
arrays support spatial selection and simultaneous localization.  Representative
tutorials and system studies cover indoor VLC, modulation and multiple access
\cite{Mapunda2020IndoorVLC,He2023VLC,Yin2016NOMAVLC}.  At the detector level,
integrating SPAD receivers and OFDM-oriented models demonstrate that dead time,
 pixel number, and digital combining determine the useful dynamic range
 \cite{Fisher2013IntegratingReceiver,Huang2023SPADOFDM}.  Indoor systems normally
 favor passively cooled silicon devices and low electrical power over the last
 fraction of a decibel in sensitivity.
Transmitter and deployment baselines include industrial, micro-LED, organic,
array-OFDM, equalized, nonlinear, and rolling-shutter VLC systems
\cite{Lit0178,Lit0250}.

\subsubsection{Terrestrial FSO links}
Terrestrial FSO links connect buildings, towers, temporary sites, or ground
terminals over hundreds of meters to tens of kilometers.  The narrow beam offers
spatial reuse and a high link rate, but fog and low cloud can dominate the link
margin; turbulence adds scintillation, beam wander, and wavefront distortion.
Daylight operation makes optical bandwidth and field of view (FOV) first-order
design variables.  Narrowband filters suppress background without changing
geometric loss \cite{Shtyrkova2019Narrowband}, and photon-counting sequence and
multiple-symbol detectors exploit temporal statistics when individual symbols
are unreliable \cite{Chatzidiamantis2010MLSD,Riediger2007MSD}.  Alternative
detection rules, learned boundaries, and multiuser interference cancellation
illustrate the broader receiver-design space
\cite{Gurugopinath2016Detection,Arya2022MLBoundaries,Zhou2013OCDMA}.  A ground
 terminal can usually accommodate a large aperture, precision mount, active
 wavefront control, and cooling, but availability under dense fog remains a
 network-level rather than a receiver-only problem.
Adaptation, diversity, relaying, coherent OFDM, space--time PPM, daylight tests,
and phased arrays illustrate the terrestrial design space
\cite{Lit0207,Lit0258,Lit0272}.

\subsubsection{UAV, aircraft, and satellite links}
Mobile aerial and orbital links combine large geometric loss with rapidly
changing range, Doppler, line of sight, and pointing geometry.  Platform jitter
and angular rate can be more important than atmospheric scintillation, especially
when most of the path lies above the turbulent boundary layer.  The receiver must
close acquisition, tracking, and communication loops without violating stringent
size, weight, and power (SWaP) limits.  Modular SNSPD ground receivers and
multi-pixel arrays show how parallel detector channels increase both maximum
count rate and outage tolerance \cite{Vyhnalek2018ModularSNSPD,Hao2024SNSPDArray}.
Photon-counting array bounds quantify the coupling between received photons and
angle-estimation accuracy \cite{Bashir2021BeamTracking}; CMOS SPAD imagers further
demonstrate that high-rate data recovery and fast spot tracking can share a focal
plane \cite{Chen2024SPADReceiver}.  For space-to-ground service, laboratory link
 emulators and field-oriented studies are essential because detector blocking,
 background, turbulence, and pointing are strongly coupled
 \cite{Marsden2024GroundReceiver,Pachedjieva2018SpaceAccuracy}.
UAV OFDM, integrated sensing, angle diversity, and optical roadmaps support
joint beam and waveform adaptation \cite{Lit0249,Lit0281}.

\subsubsection{Space-to-ground and deep-space laser communication}
Deep-space links are the canonical photon-starved case.  Aperture-to-range ratio,
transmit-beam divergence, and ephemeris error produce an enormous geometric-loss
budget, while the Earth terminal still encounters atmosphere and sky background.
High-order pulse-position modulation (PPM), coding, and low-noise time-resolved
detection are therefore used to approach high bits-per-photon efficiency.  The
system landscape and its scaling are reviewed in \cite{Wu2017DeepSpace}, and
near-single-photon-per-bit demonstrations and information-theoretic analyses
establish the sensitivity target
\cite{Caplan2006PhotonEfficient,Dolinar2012UltimateLimits,Banaszek2020QuantumLimits}.

The Lunar Laser Communication Demonstration (LLCD) showed that a fiber-coupled
SNSPD receiver can be engineered into an operational ground terminal
\cite{Grein2015LLCDReceiver,Grein2015Moon}.  Current receiver developments span
high-dynamic-range ground stations, photon-counting cameras, and coherent
combination concepts \cite{Biswas2024DeepSpacePhoton,Downey2024GroundReceiver,Licht2025PhotonCountingCamera,Mikos2026CoherentCombination}.  These systems must
also address detector saturation and focal-plane centroid bias
\cite{Srinivasan2018Centroiding}, and performance must be stated against both
photon efficiency and achievable coded throughput
\cite{Boroson2018PerformanceLimits,Caplan2024PhotonCounting}.  Device options
 include conventional SPADs, negative-avalanche-feedback devices, and SNSPDs
 \cite{Farr2005Detectors,Farr2009NAF}; selection depends as much on count-rate
 linearity and timing as on peak detection efficiency.
Scaling, camera, fiber, receiver-algorithm, PIE, turbulent-PPM, and satellite
studies document the transition to complete ground receivers
\cite{Lit0024,Lit0043,Lit0061,Lit0172,Lit0195}.

\subsubsection{Underwater wireless optical communication}
Underwater OWC commonly uses the blue--green transmission window.  Absorption
sets an exponential power penalty, scattering broadens the angular and temporal
response, and bubbles, suspended particles, and water motion make the channel
intermittent.  Photon-counting experiments have demonstrated both extreme
attenuation-length operation with high photon efficiency and gigabit-per-second
shorter links \cite{Hu2018UWOC256PPM,Zhang2020SiPMUWOC}.  SPAD systems extend
reach, while SiPM and single-SPAD OFDM receivers trade pixel dead time against
bandwidth and dynamic range
\cite{Chen2020LongDistanceUWOC,Essalih2020SiPMUWOC,Lyu2022SPADUWOC}.  Spatial
diversity and multipulse PPM reduce fading sensitivity
\cite{Peppas2017UWOCSpatialDiversity}.  For air--water links the refractive
interface and surface waves add beam displacement and angular spreading
\cite{Sun2025AirWater}.  Packaging, biofouling, pressure resistance, and the
 difficulty of rejecting solar background near the surface often dominate the
 choice of aperture and FOV; cryogenic cooling is rarely acceptable on small
 subsea platforms.
Underwater variants span multiplexing, multi-pixel counting, application
traffic, duplex operation, PPM/OFDM, aperture scaling, and moving interfaces;
reported reach must distinguish physical distance from attenuation length
\cite{Lit0009,Lit0013,Lit0039,Lit0040,Lit0042,Lit0054,Lit0071,Lit0085,Lit0103,Lit0157,Lit0173,Lit0180,Lit0187,Lit0189,Lit0192,Lit0193,Lit0197,Lit0225}.

\subsubsection{Non-line-of-sight scattering communication}
Non-line-of-sight (NLOS) links intentionally collect photons scattered by the
atmosphere or another diffuse medium.  Solar-blind ultraviolet (UV) wavelengths
can reduce daytime background, but the useful signal experiences the product of
two spreading legs and a small scattering probability.  Large FOVs increase the
captured signal and the background simultaneously.  Photon-counting count-and-
forward relays illustrate communication over such Poisson channels
\cite{Gong2015NLOS}, while dual-receiver UV systems show that the spatial count
 distribution can also support positioning \cite{Wang2023NLOSPositioning,Lit0317}.
 Unlike a narrow-beam LOS terminal, an NLOS receiver may relax fine pointing but
 requires careful overlap-volume geometry, low dark counts, and long integration.
 Nonideal-count scattering models, explicit scattering-channel estimation, and
 underwater NLOS network analysis quantify the resulting geometry--background
 tradeoff \cite{Lit0120,Lit0136,Lit0383}.

\subsubsection{Integrated communication and sensing}
The same time tags and spatial count maps used for communication contain range,
angle, and channel-state information.  Two-way photon-counting optical links can
support precision time transfer \cite{Trojanek2018TimeTransfer}; focal-plane
arrays can estimate angle while receiving data \cite{Bashir2021BeamTracking,Chen2024SPADReceiver}; and UV count maps can provide NLOS localization
\cite{Wang2023NLOSPositioning}.  This integrated communication-and-sensing mode
is attractive when aperture, detector, and pointing resources are scarce.  It
also exposes a resource conflict: narrow temporal gates and narrow FOVs suppress
background, whereas acquisition and sensing favor wider search windows.  A
 joint scheduler should therefore allocate detector pixels, gates, and pilot
 symbols between discovery, tracking, ranging, and payload decoding.
Related studies connect sensing with energy harvesting, optical antennas, and
flexible receivers \cite{Lit0249,Lit0267,Lit0268,Lit0276}.

\subsection{How the Scenarios Differ}

Three comparisons deserve emphasis.  First, wavelength is a system variable:
source efficiency, eye safety, atmospheric or water transmission, filter
technology, and detector efficiency all enter the decision.  Second, geometric
loss cannot be repaired by detector sensitivity alone; aperture, divergence,
pointing, and coupling efficiency multiply before the photon reaches the active
area.  Third, background rejection is joint spectral--spatial--temporal
filtering.  A narrower optical passband, smaller accepted etendue, and tighter
TDC gate lower $\bar n_{\mathrm{b}}$, but each can lose signal under wavelength
drift, turbulence, pointing error, or clock uncertainty.  The appropriate
receiver thus maximizes coded information rate and availability, not merely
detector efficiency.

\subsection{Photon-Counting Receiver Architectures}

\subsubsection{Free-space coupling}
The simplest architecture images the receive aperture directly onto a
single-pixel detector or focal-plane array.  It avoids fiber-coupling loss,
accepts a comparatively large etendue, and is robust to residual wavefront
aberration.  Its liabilities are detector-area capacitance for SPADs, the need
for a sufficiently large active area, and weaker spatial rejection of
background.  A single pixel is attractive when a separate tracking sensor holds
the spot on the detector and the required count rate remains below the dead-time
limit.  Parallel or multi-channel detectors extend dynamic range and tolerate
partial blockage \cite{Du2018MultiChannel,Fisher2013IntegratingReceiver}.

\subsubsection{Multimode-fiber coupling}
A multimode fiber (MMF) transports collected light from the telescope focus to
a mechanically and thermally isolated detector package.  Its larger core and
numerical aperture relax alignment and match the etendue of a turbulence-broadened
spot, which made fiber-coupled SNSPD reception practical in LLCD-class ground
terminals \cite{Grein2015LLCDReceiver,Vyhnalek2019FewMode}.  The design must
budget focal-ratio mismatch, connector loss, modal noise, bend sensitivity, and
modal dispersion.  Increasing accepted modes improves signal capture but also
admits more spatial background; the fiber is therefore part of the optical
filter, not a lossless cable.

\subsubsection{Single-mode fiber with adaptive optics}
A single-mode fiber (SMF) is an exceptionally strong spatial filter and interfaces
cleanly to integrated photonics or a small detector.  Without correction,
however, atmospheric phase errors and beam wander make coupling intermittent.
Adaptive optics (AO), tip--tilt stabilization, and precise non-common-path
calibration are required to convert telescope-aperture gain into SMF coupling
gain.  This option is most compelling at a capable ground terminal where reduced
background and photonic integration justify the actuator, wavefront-sensor, and
control complexity.  Its performance should be evaluated using the distribution
of instantaneous coupling efficiency, not only its diffraction-limited peak.

\subsubsection{Photonic lanterns}
A photonic lantern adiabatically maps a multimode input into several single-mode
outputs.  It therefore occupies the middle ground between robust MMF collection
and SMF-compatible detection.  Each output can feed a detector pixel, an
integrated filter, or a diversity combiner, while the vector of output counts can
carry wavefront or pointing information.  Measurements under emulated
atmospheric conditions show its relevance to low-Earth-orbit ground receivers
\cite{Tedder2020PhotonicLantern}.  The price is insertion loss, multiple detector
and TDC channels, channel-to-channel calibration, and a combining rule that
 remains robust when the modal distribution changes.
Hybrid engines and concentrators offer alternative area scaling
\cite{Lit0202,Lit0218}.

\subsubsection{Single-pixel, array, and imaging receivers}
A single-pixel SPAD or SNSPD minimizes readout complexity and can deliver the
best per-channel timing performance, but its dead time caps the count rate and it
cannot distinguish a displaced spot from a power fade.  SPAD/SNSPD arrays divide
the flux across pixels, raising aggregate count-rate capability and providing
spatial observability; they also introduce fill-factor loss, pixel nonuniformity,
crosstalk, and a large event-readout load.  High-capacity space-to-ground work
uses multi-pixel SNSPDs for precisely this sensitivity--dynamic-range trade
\cite{Hao2024SNSPDArray,Caplan2024PhotonCounting}.  Imaging receivers add a lens
or telescope mapping so that selected pixels implement an adaptive FOV.  They
can reject off-axis background and estimate centroid, but detector blocking and
pile-up can bias the centroid at high flux \cite{Srinivasan2018Centroiding}.
 Thus, pixel count is not a performance metric by itself; active area, fill
 factor, photon-detection efficiency, timing jitter, per-pixel dead time, maximum
 aggregate event rate, and readout bandwidth must be reported together.
Array statistics and alternative transmitter/collector geometries support this
co-design \cite{Lit0079,Lit0113,Lit0125,Lit0128,Lit0130,Lit0207,Lit0236,Lit0243}.

\subsubsection{TDC, FPGA, synchronization, and decoding}
The detector output is an event stream rather than a conventional analog
waveform.  A TDC records photon arrival times or accumulates gated histograms;
an FPGA then performs event validation, binning, frame and slot synchronization,
background estimation, demodulation, soft-metric generation, and FEC interfacing.
NASA's real-time photon-counting receiver is an instructive engineering case:
the detector, time-to-digital conversion, and FPGA processing were integrated to
recover high-photon-efficiency optical waveforms in real time
\cite{Vyhnalek2019Realtime}.  This partition is also useful for terrestrial and
underwater links.  The critical interfaces are detector pulse width versus TDC
double-hit resolution, aggregate event rate versus I/O bandwidth, clock stability
versus gate width, and buffer depth versus fading bursts.  Receiver algorithms
 must use a dead-time-aware likelihood when the flux is high; a nominal Poisson
 model otherwise overestimates both signal and information rate.
Array acquisition and time tagging expose digital-interface limits
\cite{Lit0070,Lit0075,Lit0099,Lit0260}.

\subsubsection{Joint detection, communication, and tracking}
An array or imaging focal plane enables a unified terminal: photon locations
produce centroid or angle estimates, time tags carry synchronization and ranging,
and the same events feed payload decoding.  The estimated angle drives a fast
steering mirror or AO stage, while clock estimates return to the TDC gate.
Joint processing reduces optical
duplication and non-common-path error, but a bright acquisition mode can saturate
a detector optimized for weak communication.  Practical designs therefore need
multiple gain or gating modes, protected search patterns, count-rate monitors,
and explicit handover thresholds from acquisition to fine tracking.  Array
tracking limits \cite{Bashir2021BeamTracking}, photon-counting camera development
\cite{Licht2025PhotonCountingCamera}, and real-time ground-receiver experiments
 \cite{Vyhnalek2019Realtime,Marsden2024GroundReceiver} collectively support this
 closed-loop architecture.
Daylight, energy-sharing, phased-array, and UAV studies extend this architecture
\cite{Lit0258,Lit0267,Lit0272,Lit0281}.

In summary, architecture selection should proceed from the scenario to the
photon budget and only then to a detector.  The designer first fixes wavelength,
aperture, FOV, availability, and PAT error; next chooses spectral and spatial
coupling; then allocates count-rate and timing requirements across pixels and
TDC channels; and finally co-designs synchronization, modulation, and coding.
This sequence prevents a high-efficiency detector from being paired with an
optical front end that loses the signal, admits excessive background, or cannot
maintain alignment.

\section{Single-Photon Detector Technologies, Performance Metrics, and Engineering Selection}
\label{sec:detector_technologies}

The receiver architectures in Section~\ref{sec:scenarios_architectures} produce
either a focused free-space spot or one or more fiber outputs.  The detector must
convert that optical field into a sequence of electrical events without erasing
the time, photon-number, or spatial information required by the modem.  This
section therefore separates four questions that are sometimes mixed together:
how a photon becomes a registered event; how the principal detector families
implement that conversion; how performance and nonidealities should be measured;
and how a technology should be selected and integrated.  Sections III-A and
III-B address the first two questions.  Sections III-C and III-D use the same
notation to connect device behavior to communication-level engineering choices.

\subsection{From an Incident Photon to a Registered Event}
\label{subsec:detector_principles}

A photon at wavelength $\lambda$ carries energy
\begin{equation}
  E_{\gamma}=h\nu=\frac{hc}{\lambda}.
  \label{eq:photon_energy}
\end{equation}
Detection requires absorption of this energy and conversion of the resulting
excitation into an electrical signal that crosses a discriminator threshold.
It is useful to factor the probability of this chain as
\begin{equation}
  \eta_{\mathrm{det}}(\lambda)=
  \eta_{\mathrm{cpl}}\,f_{\mathrm{act}}\,
  \eta_{\mathrm{abs}}(\lambda)\,p_{\mathrm{conv}},
  \label{eq:detection_efficiency_factorization}
\end{equation}
where $\eta_{\mathrm{cpl}}$ is the coupling efficiency from the receiver optical
port to the sensitive area, $f_{\mathrm{act}}$ represents fill factor or other
geometrical coverage, $\eta_{\mathrm{abs}}$ is the probability of absorption in
the active material, and $p_{\mathrm{conv}}$ is the conditional probability that
an absorbed photon produces a registered electrical event.  The factors are not
always independently measurable, so authors variously report quantum efficiency,
device photon-detection efficiency (PDE), or end-to-end system detection
efficiency.  A tutorial comparison must preserve this distinction; inserting a
device PDE into a link budget that already includes coupling or fill-factor loss
double counts a gain.

For an ideal binary detector observed over a gate of duration $T_{\mathrm g}$,
independent Poisson arrivals give the Bernoulli click probability
\begin{equation}
 p_{\mathrm{click}}=1-\exp\!\left[-\left(
 \eta_{\mathrm{det}}\bar n_{\mathrm{s}}+\bar n_{\mathrm{b}}
 +r_{\mathrm d}T_{\mathrm g}\right)\right],
 \label{eq:ideal_click_probability}
\end{equation}
where $\bar n_{\mathrm{s}}$ is the incident signal-photon number, $\bar
n_{\mathrm{b}}$ is the mean registered background count, and $r_{\mathrm d}$ is
the dark-count rate.  This form is widely used as the front end of
photon-counting communication models \cite{Karp1970PhotonCounting,Yiannopoulos2019ErrorProbability}, but it is an interface model rather than a
complete detector model: it ignores recovery, afterpulsing, crosstalk, readout
contention, and threshold errors.  Nevertheless, it exposes a key point.  A
photon counter returns a random point process
$i_{\mathrm e}(t)=\sum_k a_k\delta(t-t_k)$ or a binned count, rather than the
continuous photocurrent assumed by a conventional linear receiver.  Its output
can retain arrival time $t_k$, pixel address, and sometimes pulse amplitude, but
only if the readout preserves them.  Classical photon-counting detector surveys
and communication analyses make the same separation between optical conversion,
event formation, and receiver processing \cite{Farr2005Detectors,Karp1970PhotonCounting,Yiannopoulos2019ErrorProbability}.

\subsection{Detector Technologies}
\label{subsec:detector_technologies}

No detector family simultaneously maximizes active area, system detection
efficiency, count rate, timing precision, spectral coverage, operating
temperature, and integration density.  The useful comparison is therefore not
``which detector is best,'' but which physical transduction mechanism and
readout observable preserve the information required by a particular OWC link.
Table~\ref{tab:detector_technology_comparison} summarizes this system-level
view.  The entries are deliberately qualitative because numerical performance
depends strongly on wavelength, bias, temperature, coupling, illuminated area,
and readout configuration.

\begin{table*}[!t]
  \caption{System-Level Comparison of Photon-Counting Detector Technologies for OWC}
  \label{tab:detector_technology_comparison}
  \centering
  \scriptsize
  \setlength{\tabcolsep}{3pt}
  \renewcommand{\arraystretch}{1.18}
  \begin{tabularx}{\textwidth}{p{0.105\textwidth}p{0.175\textwidth}p{0.135\textwidth}YYY}
    \toprule
    \textbf{Technology} & \textbf{Physical event} & \textbf{Native observable} &
    \textbf{Principal strengths} & \textbf{Principal constraints} &
    \textbf{Representative OWC fit} \\
    \midrule
    SPAD & Geiger avalanche in one reverse-biased junction & Binary click; time tag when paired with a TDC & CMOS compatibility, gating, low-voltage digital interface, visible-to-SWIR material options & Dead time, afterpulsing, binary saturation, temperature-sensitive excess bias & Compact VLC, underwater links, and time-resolved receivers \cite{Liu2025NovelCMOSSPAD,Det0064,Det0092} \\
    SiPM & Parallel firing of many passively quenched SPAD microcells & Summed current, charge, or pulse amplitude & Millimeter-scale area, high gain, approximate photon-number resolution, simple analog summation & Microcell occupancy, recharge, crosstalk, capacitance, and area-scaled dark rate & Wide-FOV and alignment-tolerant indoor or underwater terminals \cite{Li2025SiPMModes,Det0020,Det0108} \\
    CMOS SPAD array & Addressed Geiger events distributed across integrated pixels & Count, pixel address, histogram, or per-event time tag & Imaging, time gating, spatial filtering, tracking, and on-chip processing & Fill-factor loss, nonuniformity, hot pixels, TDC sharing, arbitration, and output-bandwidth limits & High-speed links and joint communication, ranging, and tracking \cite{Lit0357,Chen2024SPADReceiver,Det0384,Det0390} \\
    SNSPD & Photon-induced resistive transition in a current-biased superconducting nanowire & Fast electrical pulse; time tag; array address if instrumented & High telecom-band efficiency, low dark counts, small jitter, and scalable parallelism & Cryogenic cooling, small active area, optical coupling, RF readout, and recovery inductance & Ground terminals, space-to-ground reception, and deep-space optical links \cite{Hao2024SNSPDArray,Wan2024DaytimeSNSPD,Det0086} \\
    PMT/MCP-PMT & Photoemission followed by dynode or microchannel electron multiplication & Analog pulse amplitude and arrival time & Large photosensitive area, high gain, mature fast timing, broad free-space acceptance & High voltage, bulk, ageing, gain spread, and magnetic sensitivity & Scattering, NLOS, underwater, and legacy photon-counting receivers \cite{Det0077,Det0138} \\
    Upconversion SPD & Nonlinear sum-frequency conversion followed by visible-band photon detection & Click or time tag from the secondary detector & Access to infrared bands using efficient silicon detectors; spectral and modal filtering & Strong pump, conversion noise, phase matching, insertion loss, and limited angular/mode acceptance & Fiber-coupled or mode-controlled infrared PPM links \cite{Det0271,Det0049} \\
    Emerging/hybrid & Calorimetric, resonant, heterogeneous, or switched multi-detector conversion & Energy, number, time, position, or adaptively selected analog/click output & New observables and operating-point adaptability & Lower maturity, specialized refrigeration or fabrication, and incomplete communication benchmarks & Research links requiring energy resolution, extreme sensitivity, or large dynamic range \cite{Det0230,Det0037} \\
    \bottomrule
  \end{tabularx}
\end{table*}

\subsubsection{Single-photon avalanche diodes}

A single-photon avalanche diode (SPAD) is a $p$--$n$ or $p$--$i$--$n$ junction
operated in Geiger mode at a reverse bias $V_{\mathrm{bias}}$ above its breakdown
voltage $V_{\mathrm{BD}}$.  The excess bias
\begin{equation}
  V_{\mathrm{ex}}=V_{\mathrm{bias}}-V_{\mathrm{BD}}>0
  \label{eq:spad_excess_bias}
\end{equation}
creates a field in which a single photogenerated carrier can initiate impact
ionization.  Because the resulting avalanche is self-sustaining, the SPAD does
not measure the initiating photon's energy or directly distinguish one photon
from several photons arriving in the same unresolved interval.  Instead, a
quenching circuit lowers the junction voltage below breakdown, stops the
avalanche, and then recharges the diode \cite{Chitnis2014SPADSystem,Fisher2017CMOSSPAD,Wang2019SPADOperation}.

Passive quenching uses a large series resistance and has a recovery time of the
order of $R_{\mathrm q}C_{\mathrm{eq}}$, whereas active quenching senses the
avalanche and forces a controlled hold-off and reset.  The latter makes the dead
time more reproducible and can support higher count rates, at the cost of pixel
area, switching power, and circuit complexity.  A fully integrated $4\times4$
receiver, for example, combined tunable quench/reset circuits with an OR tree and
reported a minimum dead time of approximately $2.5$~ns
\cite{Li2022IntegratedSPAD}; modern CMOS pixels also demonstrate that very small
single-photon timing response and few-nanosecond dead time can coexist, although
not independently of active area, excess bias, and noise
\cite{Gramuglia2021LowNoiseSPAD}.

Increasing $V_{\mathrm{ex}}$ generally raises avalanche-trigger probability and
hence PDE, but also tends to increase dark counts, afterpulsing, optical
crosstalk, and power.  Breakdown voltage is temperature dependent, so fixed-bias
operation changes the effective excess bias as the terminal temperature drifts.
Material choice sets the useful spectral range: silicon SPADs are attractive in
the visible and short-wave NIR, while InGaAs/InP structures extend Geiger-mode
detection to the $1.3$--$1.55~\mu$m bands at the price of more demanding noise,
gating, and temperature control.  For communication, the practically relevant
object is therefore the \emph{SPAD plus quench, threshold, and readout}, not the
bare junction.

The idealized response of a binary SPAD makes two consequences explicit.  In the
photon-starved region, the click probability is
approximately linear in $\eta_{\mathrm{det}}\bar n_{\mathrm s}$, so higher PDE
directly improves sensitivity.  At larger photon number, however, the binary
output approaches unity and no longer conveys proportional optical intensity.
Background and dark counts also establish a nonzero click floor.  Dead time then
adds a second saturation mechanism because a registered avalanche temporarily
removes that pixel from the available detector population
\cite{Sarbazi2018SPADStatistics,Huang2023SPADOFDM}.

The digital-like output is especially compatible with CMOS integration.  A
small SPAD array can reduce required optical power density while preserving
photon counting \cite{Chitnis2014SPADSystem}, and commercial arrays have supported
multi-gigabit OWC when their dead-time nonlinearity is handled by modulation and
equalization \cite{Huang2022FiveGbpsSPAD,Huang2023SPADOFDM}.  These results do not
make a SPAD universally more sensitive than an APD.  Comparisons must include
PDE, active area, output pulse width, fill factor, background level, and the
bandwidth or dead-time constraint; measured VLC comparisons show that each
receiver occupies a different useful optical-power regime
\cite{Zhang2018APDSPADComparison}.

SPAD experiments span indoor, underwater, FSO, and inter-satellite links
\cite{Det0162,Det0166,Det0397,Det0073,Det0064,Det0092}.  Gate control,
attenuation, sub-dead-time reception, and nonlinear compensation address
different parts of waveform-dependent recovery
\cite{Det0134,Det0062,Det0091,Det0095,Det0123,Det0167,Det0393,Det0016,Det0051,Det0053,Det0389,Det0041,Det0122}.  Timing, learned detection,
multichannel readout, and integrated arbitration add shared-resource constraints
\cite{Det0014,Det0006,Det0069,Det0156,Det0026,Det0025,Det0121,Det0164,Det0385,Det0390,Det0384}.  VLC and underwater results confirm that detector and
DSP choices are coupled \cite{Det0110,Det0388,Det0055,Det0066,Det0044,Lit0387,Lit0400}.

\subsubsection{Silicon photomultipliers}

A silicon photomultiplier (SiPM) is a dense array of SPAD microcells connected to
a common output, with each cell usually passively quenched.  If a fired cell
delivers charge $Q_{\mathrm{cell}}\simeq C_{\mathrm{cell}}V_{\mathrm{ex}}$, the
integrated output charge is approximately $M Q_{\mathrm{cell}}$ when $M$ cells
fire.  Thus, unlike one binary SPAD, a SiPM can provide limited photon-number
resolution from aggregate charge or pulse amplitude.  Neglecting recovery,
afterpulsing, crosstalk, and nonuniform illumination, uniform illumination of
$N_{\mathrm{cell}}$ cells gives the useful occupancy model
\begin{equation}
 \mathbb{E}[M]\simeq N_{\mathrm{cell}}
 \left[1-\exp\!\left(-\frac{\eta_{\mathrm{PDE}}\bar n}
 {N_{\mathrm{cell}}}\right)\right].
 \label{eq:sipm_occupancy}
\end{equation}
The low-flux slope is proportional to PDE, but the response compresses when
multiple photons compete for already fired cells.  Cell recharge allows refiring
for waveforms longer than the recovery time, so a communication-grade dynamic
model must include both spatial occupancy and temporal recovery
\cite{Essalih2020SiPMUWOC,He2022SiPMSelection}.  Increasing
$N_{\mathrm{cell}}$ postpones occupancy saturation without changing the
low-flux slope.  The apparent photon-number resolution is therefore conditional:
it is useful only while the pulse-charge distributions remain separable and the
available-cell population is known.

SiPMs offer millimetre-scale collection area, internal gain, and simple analog
summation without one TDC per microcell.  Their communication limitations include
cell capacitance and output pulse width, correlated afterpulses, optical
crosstalk, temperature-dependent gain, finite cell count, and increased dark
rate with area.  Device selection is consequently a joint choice of microcell
size, cell count, PDE, active area, terminal capacitance, and pulse shape rather
than a search for the largest quoted PDE.  This co-design has enabled both
long-reach underwater links \cite{Zhang2020SiPMUWOC,Essalih2020SiPMUWOC} and an
$8.2$-Gb/s NIR OWC experiment using a fast commercial SiPM
\cite{Ali2023SiPM82Gbps}.

Large-area SiPM studies use diversity, positioning, tracking, NLOS, and IoT
architectures \cite{Det0020,Det0380,Det0382,Det0378,Det0108,Det0056,Det0379}.
Dual-readout, MIMO, equalization, and multistate operation must track cell
availability and analog bandwidth \cite{Det0398,Det0027,Det0067,Det0158,Det0036,Li2025SiPMModes,Det0093,Det0023,Det0396}.  Cell pitch, area, bias,
temperature, and background therefore accompany PDE
\cite{Zhang2022SiPMGuideline,Det0142,Det0143,Det0109,Det0080,Det0386}.

\subsubsection{CMOS SPAD arrays}

A CMOS SPAD array uses similar Geiger microcells but exposes their events through
digital logic, counters, arbiters, pixel addresses, or on-chip TDCs rather than
only summing all avalanche charge.  It can therefore perform time gating,
imaging, background rejection, centroiding, and per-pixel calibration.  The price
is reduced optical fill factor and a possible electronic bottleneck: many photons
may arrive simultaneously, while a shared bus, OR tree, counter, or TDC can
accept only a finite event rate.  SiPM and digital SPAD-array results must not be
merged into one category solely because both contain SPAD microcells; their
observable data and saturation mechanisms are different.

Array architecture is consequently part of the detector specification.  A
global counter minimizes area but discards position and may serialize coincident
events; per-pixel counters preserve spatial counts but consume fill factor and
power; time-to-digital conversion adds timing information at still higher data
volume.  Column-wise decoding and fast spot tracking demonstrate the benefit of
retaining spatial structure \cite{Chen2024SPADReceiver}, while integrated and
time-controlled arrays show how local gating can suppress intersymbol
interference \cite{Li2022IntegratedSPAD,Liu2023TimeControlledSPAD,Liu2025NovelCMOSSPAD}.  APD/SPAD comparisons must therefore state whether a
single junction, summed array, or addressable imager is being evaluated
\cite{Det0117}.

\subsubsection{Superconducting nanowire single-photon detectors}

An SNSPD is a narrow superconducting strip, commonly patterned as a meander and
biased at a current $I_{\mathrm b}$ below its switching or critical current.  An
absorbed photon creates nonequilibrium quasiparticles and locally suppresses the
superconducting order parameter.  Hotspot-assisted and vortex-mediated models
describe how this perturbation develops into a resistive region; the microscopic
boundary between these pictures depends on material, geometry, bias, photon
energy, and temperature \cite{Zhang2021SNSPDMechanism}.  Circuit current is then
diverted into the load, producing a voltage pulse.  After the resistive region
cools, the current returns with a characteristic electrical recovery on the order
of
\begin{equation}
  \tau_{\mathrm{reset}}\sim\frac{L_{\mathrm k}}{R_{\mathrm L}},
  \label{eq:snspd_reset}
\end{equation}
where $L_{\mathrm k}$ is the nanowire kinetic inductance and $R_{\mathrm L}$ is
the effective load seen during reset.  This compact expression is useful for
intuition but does not capture latching, electrothermal feedback, readout
bandwidth, or a distributed multi-pixel geometry.  With a first-order return
$I(t)/I_{\mathrm b}=1-\exp(-t/\tau_{\mathrm{reset}})$, the detector reaches
90\% of its bias after $t_{90}=\tau_{\mathrm{reset}}\ln 10$; thus even a short
electrical time constant occupies several time constants before the detection
efficiency is nearly restored.

For SNSPDs, system detection efficiency is often decomposed into optical coupling,
absorptance, and internal detection efficiency.  Optical stacks and cavities can
raise absorptance, a sufficiently high bias may produce a plateau in internal
efficiency, and careful packaging determines whether that device performance is
retained at the receiver port.  Free-space coupling at $1.55~\mu$m has been
demonstrated with cryogenic optical trains \cite{Bellei2016FreeSpaceSNSPD}, while
fiber-coupled SNSPD receivers enabled the Lunar Laser Communication Demonstration
\cite{Grein2014FiberSNSPD,Grein2015LLCDReceiver}.  Materials and geometry can also
move operation toward less severe cooling; NbTiN devices operating above $4$~K,
for example, illustrate that efficiency, jitter, bias margin, and temperature
must be reported together \cite{Gourgues2019HighTempSNSPD}.

The central communications advantage is the combination of high efficiency at
telecom wavelengths, low dark-count rate, small timing jitter, and high count-rate
potential.  The engineering cost is a cryostat, thermal interfaces, RF readout,
small sensitive area, and possible coupling sensitivity.  Parallel pixels trade
additional cryogenic wiring and readout load for aggregate count rate, spatial
information, and fault tolerance.  Recent arrays combine near-unity efficiency,
gigacount-per-second aggregate operation, photon-number discrimination, and
coarse position sensing in a space-to-ground link experiment
\cite{Hao2024SNSPDArray}; detector and receiver reviews emphasize that the usable
system count rate depends as much on coupling and readout as on nanowire recovery
\cite{Guardiani2024SNSPDLaserCom,Vyhnalek2018ModularSNSPD}.

Packaged SNSPD arrays emphasize modularity, aggregate rate, background,
real-time time tags, coupling, and cryogenic operation
\cite{Det0086,Det0072,Det0129,Det0392,Det0135,Det0010,Det0145,Det0165,Det0046,Det0050,Wan2024DaytimeSNSPD,Srinivasan2023DSOCReceiver}.

\subsubsection{Photomultiplier and microchannel-plate detectors}

Photomultiplier tubes (PMTs) convert photons at a photocathode and multiply the
photoelectron through dynodes; microchannel-plate PMTs replace the dynode chain
with microscopic multiplication channels.  Their large photosensitive areas,
high gain, and fast variants remain useful in scattering, underwater, and legacy
deep-space receivers, particularly when free-space collection is more important
than CMOS integration.  High voltage, magnetic sensitivity, ageing, size, and
gain statistics are corresponding system considerations
\cite{Farr2005Detectors,Gong2015NLOS}.  Arrayed PMTs can supply spatial diversity
without semiconductor fill-factor loss, as demonstrated in photon-counting MIMO
underwater links \cite{Det0077}; PMT-based scattering-channel estimation also
shows the value of a large collection area in NLOS geometries \cite{Det0138}.
Unlike an ideal click detector, however, a PMT produces a stochastic analog
pulse whose gain distribution, discriminator threshold, and pileup affect the
observed count.  A reproducible comparison must therefore state whether pulses
were counted, integrated, or processed as a continuous waveform.

\subsubsection{Upconversion single-photon detectors}

An upconversion receiver translates an infrared signal to a wavelength at which
a mature silicon detector can be used.  In a periodically poled nonlinear
crystal, a signal photon at angular frequency $\omega_{\mathrm s}$ and a strong
pump at $\omega_{\mathrm p}$ generate a sum-frequency photon at
$\omega_{\mathrm u}=\omega_{\mathrm s}+\omega_{\mathrm p}$.  Spectral filters
then reject the pump and parasitic light before the translated photon reaches a
silicon SPAD or related counter.  This architecture can combine infrared link
wavelengths with efficient, low-jitter visible-band detection, and has been used
in photon-counting PPM communication receivers \cite{Det0049}.

The conversion stage is also a selective optical preprocessor.  Its phase-
matching bandwidth and spatial-mode acceptance can reject out-of-band or poorly
matched background, while waveguide implementations can provide high internal
conversion efficiency.  The cost is pump power, insertion loss, alignment,
conversion-induced noise, and a restricted optical mode set.  Mid-infrared
photon counting and number resolution demonstrate the widening spectral reach
of this approach \cite{Det0271}, but a communication link budget must use the
end-to-end detection efficiency after nonlinear conversion and filtering, not
the silicon detector PDE alone.

\subsubsection{Emerging and hybrid detectors}

Linear-mode APDs remain an important boundary technology.  They provide analog
avalanche gain and potentially large bandwidth, but multiplication noise and
front-end noise distinguish them from Geiger-mode event counters
\cite{Zhang2018APDSPADComparison}.  Recent deep-space uplink analyses retain APDs
as competitive candidates when the received flux, bandwidth, and implementation
constraints do not require discrete-event operation \cite{Det0230}.  Hybrid
PD/SPAD receivers extend this idea by switching or combining analog and
photon-counting paths to cover a wider dynamic range \cite{Det0062}.

Transition-edge sensors and related cryogenic calorimeters can resolve photon
energy and photon number with exceptional sensitivity, but their recovery time,
sub-kelvin refrigeration, and multiplexed readout generally make them less suited
to high-rate OWC than SNSPDs.  Resonator-based detectors, heterogeneous
semiconductor absorbers, and alternative superconducting geometries may
eventually trade energy resolution, wavelength range, or multiplexability
against speed.  At the system level, new optical interfaces can be equally
important: a recent coiled-fiber receiver, for example, converts alignment and
collection constraints without changing the underlying click detector
\cite{Det0037}.

The terminology should therefore follow the receiver observable.  A
\emph{click detector} reports whether an event occurred, a \emph{time-tagging
detector} reports $t_k$, an \emph{imaging detector} also reports position, and a
\emph{photon-number-resolving detector} returns a calibrated estimate of photon
number.  Declaring that observable prevents later channel and decoder models
from assuming information that the hardware never produces.

The technology choice emerging from this subsection is qualitative but already
useful.  CMOS SPADs favor compact integration, time gating, and imaging; SiPMs
favor large area and inexpensive analog summation; and SNSPDs favor the lowest
noise and strongest telecom-wavelength timing performance when cryogenic ground
infrastructure is acceptable.  Section III-C turns these statements into
measurable quantities, while Section III-D combines nonideal event models with
engineering selection and maps the metrics back to the scenarios and receiver
ports defined in Section~\ref{sec:scenarios_architectures}.

\subsection{Key Performance Metrics and Characterization Methods}
\label{subsec:detector_metrics}

A detector data sheet is meaningful for an OWC link only when each number is
attached to its measurement plane and operating condition.  Intrinsic absorption
or avalanche-trigger probability, packaged-device PDE, and receiver-level system
detection efficiency are different quantities; similarly, a pixel dead time is
not automatically the maximum count rate of an array.  Detector and receiver
studies repeatedly show that wavelength, bias, temperature, illuminated area,
 optical coupling, discriminator threshold, gate width, and readout architecture
 must accompany the headline metric \cite{Farr2005Detectors,Huang2023SixG,Guardiani2024SNSPDLaserCom}.  Array and receiver demonstrations reinforce the
 same reporting requirement \cite{Hao2024SNSPDArray,Liu2025NovelCMOSSPAD,Caplan2024PhotonCounting}.  This subsection establishes
a common reporting framework.  It defines what should be measured and how the
measurement should be qualified; Section III-D will use these quantities in
communication-channel models.

\subsubsection{Detection efficiency and spectral response}

For a calibrated incident photon flux $\Phi_{\mathrm{in}}$ and sufficiently low
occupancy, a directly measurable efficiency is
\begin{equation}
  \eta_{\mathrm{meas}}(\lambda)=
  \frac{R_{\mathrm{on}}-R_{\mathrm{off}}}{\Phi_{\mathrm{in}}},
  \label{eq:measured_efficiency}
\end{equation}
where $R_{\mathrm{on}}$ and $R_{\mathrm{off}}$ are the illuminated and dark count
rates measured with identical bias, temperature, threshold, and timing settings.
Equation~\eqref{eq:measured_efficiency} becomes a device PDE when
$\Phi_{\mathrm{in}}$ is referenced to the active area or package input, and a
system detection efficiency (SDE) when it is referenced to the receiver optical
port.  For pulsed illumination, the incident flux should be reported as photons
per pulse and the count probability should be corrected for the probability of
two or more incident photons.  A calibrated attenuator, power meter, repetition
rate, spot size, and uncertainty budget are therefore part of the result, rather
than laboratory detail that can be omitted.

 In a SPAD, PDE combines absorption, carrier transport, avalanche-trigger
 probability, and geometrical fill factor.  It changes with wavelength, excess
 bias, junction design, temperature, and the definition of active area
 \cite{Chitnis2014SPADSystem,Fisher2017CMOSSPAD,Wang2019SPADOperation}.
 Low-noise and time-controlled devices show how this dependence changes with
 operating mode \cite{Gramuglia2021LowNoiseSPAD,Liu2023TimeControlledSPAD,Liu2025NovelCMOSSPAD}.  Microlenses may improve the effective fill factor of
a CMOS array without changing the intrinsic sensitivity of its junctions.  In a
SiPM, the quoted PDE additionally depends on microcell pitch and inactive
 quenching structures; the device that maximizes PDE may therefore have fewer
 cells, greater capacitance, or a smaller linear range \cite{He2022SiPMSelection,Zhang2022SiPMGuideline,Zhang2023SiPMEqualization}.  High-rate and multi-mode
 results expose the same compromise \cite{Ali2023SiPM82Gbps,Li2025SiPMModes}.  These tradeoffs explain
why PDE, active area, fill factor, cell count, and recovery time should be listed
together.

For an SNSPD, authors commonly separate coupling efficiency, optical absorptance,
and internal detection efficiency.  A saturated internal-efficiency plateau does
not guarantee high SDE if the optical mode misses the nanowire or packaging loss
is large.  Consequently, free-space and fiber-coupled results must state the
reference plane, polarization state, fiber type, alignment procedure, and cold
optical loss \cite{Bellei2016FreeSpaceSNSPD,Grein2014FiberSNSPD,Grein2015LLCDReceiver,Vyhnalek2019FewMode}.  Bias-current sweeps and wavelength
sweeps are more informative than a single peak number because they reveal the
usable bias margin and spectral bandwidth.  Measurements at elevated cryogenic
temperature and in multi-pixel receivers further demonstrate that efficiency,
jitter, count rate, and operating temperature cannot be optimized independently
\cite{Gourgues2019HighTempSNSPD,Vyhnalek2018ModularSNSPD,Hao2024SNSPDArray}.

\subsubsection{Dark counts, background, and sensitivity}

The receiver reports counts in the absence of a transmitted symbol for two
different reasons.  The detector and its electronics generate a dark-count rate
$R_{\mathrm d}$, while admitted environmental photons generate a background
rate $R_{\mathrm b}$.  With a gate of width $T_{\mathrm g}$, the corresponding
low-occupancy mean is
\begin{equation}
  \bar n_0=(R_{\mathrm d}+R_{\mathrm b})T_{\mathrm g}.
  \label{eq:false_count_mean}
\end{equation}
For independent Poisson arrivals, the probability of at least one false event is
$P_0=1-\exp(-\bar n_0)$.  This expression shows why a DCR quoted in counts per
second cannot be interpreted without the
decision-window duration: the same $10^6$~s$^{-1}$ environment gives
$P_0\simeq10^{-5}$ in a 10-ps gate but $P_0\simeq0.63$ in a 1-$\mu$s gate.
Recent time-controlled and time-gated SPAD receivers exploit this scaling while
also accounting for the signal photons discarded outside the gate
\cite{Huang2021TimeGated,Liu2023TimeControlledSPAD,Liu2025NovelCMOSSPAD}.
The two contributions must be measured separately: $R_{\mathrm d}$ with the
optical input blocked, and $R_{\mathrm b}$ with the complete optical path exposed
to a representative scene but the transmitter disabled.  Dark counts should be
reported as both total counts per second and, for area comparisons, counts per
second per unit active area.  Gated devices also require a per-gate probability,
gate width, repetition rate, and gate position.  Poisson models are a useful
baseline \cite{Karp1970PhotonCounting,Yiannopoulos2019ErrorProbability}, but a
dark histogram should be checked for afterpulsing, burst noise, readout artifacts,
and pixel-to-pixel dispersion before independence is assumed
\cite{Sarbazi2018SPADStatistics,Fisher2013IntegratingReceiver}.

Background is a receiver property rather than a detector constant.  Its value
depends on aperture, optical throughput, filter transmission and bandwidth, FOV,
pixel selection, wavelength, pointing direction, time of day, and the radiance of
the scene.  Narrow spectral filters reduce solar background but may introduce
angle-dependent passband shift or insertion loss \cite{Shtyrkova2019Narrowband,Rogalin2021HighDynamicRange}.  Spatial filtering and time gating can be equally
important: a single-mode port rejects off-axis modes, an imaging receiver can
activate only pixels around the tracked spot, and a PPM receiver can restrict the
decision interval \cite{Srinivasan2018Centroiding,Chen2024SPADReceiver,Vyhnalek2019Realtime}.  Ground-receiver tests should therefore record background
through the installed telescope and filter instead of substituting the detector's
laboratory DCR \cite{Marsden2024GroundReceiver,Downey2024GroundReceiver,Licht2025PhotonCountingCamera,Wan2024DaytimeSNSPD}.

A frequently quoted sensitivity surrogate is the noise-equivalent power.  Under
a one-sided shot-noise convention for a free-running detector,
\begin{equation}
  \mathrm{NEP}_{\mathrm{sp}}
  =\frac{h\nu}{\eta_{\mathrm{sys}}}\sqrt{2R_{\mathrm d}},
  \label{eq:single_photon_nep}
\end{equation}
with units of $\mathrm{W}/\sqrt{\mathrm{Hz}}$.  Equation~\eqref{eq:single_photon_nep}
is useful only when its bandwidth convention, efficiency reference plane, and
dark-count definition are stated.  It does not capture background-dominated
operation, dead-time saturation, modulation, coding, or the gain noise of an
analog SiPM output.  For communications, receiver sensitivity should also be
reported at a specified BER or frame-error rate, data rate, modulation, code,
background, and synchronization condition.  Experimental comparisons between
APD, SPAD, SiPM, and SNSPD receivers confirm that a single NEP or minimum-power
number does not preserve these operating conditions
\cite{Zhang2018APDSPADComparison,Gurugopinath2016Detection,Riediger2007MSD,Banaszek2020QuantumLimits}.

\subsubsection{Timing jitter, dead time, and recovery}

Timing jitter is measured from the distribution of the delay between a reference
optical pulse and the registered electrical event.  The histogram should be
reported with its full width at half maximum (FWHM), and preferably also its rms
width and non-Gaussian tail, because equal FWHM values can conceal very different
late-event probabilities.  If approximately independent Gaussian contributions
are applicable, the measured variance can be budgeted as
\begin{equation}
  \sigma_{\mathrm{meas}}^2=
  \sigma_{\mathrm{laser}}^2+\sigma_{\mathrm{det}}^2+
  \sigma_{\mathrm{readout}}^2+\sigma_{\mathrm{TDC}}^2+
  \sigma_{\mathrm{sync}}^2,
  \label{eq:timing_budget}
\end{equation}
where the detector contribution is commonly called the single-photon timing
resolution.  A defensible measurement states the laser pulse width, reference
clock, discriminator method, TDC bin size, count level, wavelength, bias, and
temperature.  Low-noise CMOS SPAD results illustrate the coupling among threshold
crossing, excess bias, active area, dead time, and timing response
\cite{Gramuglia2021LowNoiseSPAD,Li2022IntegratedSPAD}; SNSPD receivers add the
effects of nanowire geometry, optical absorption position, bias current, RF chain,
and cryogenic packaging \cite{Hao2024SNSPDArray,Gourgues2019HighTempSNSPD}.

For a centered decision window of width $T_{\mathrm w}$ and a Gaussian timing
response with standard deviation $\sigma$, the retained fraction is
\begin{equation}
  P_{\mathrm{cap}}(T_{\mathrm w})=
  \operatorname{erf}\!\left(\frac{T_{\mathrm w}}
  {2\sqrt{2}\sigma}\right), \qquad
  \sigma=\frac{\mathrm{FWHM}}{2\sqrt{2\ln2}}.
  \label{eq:timing_capture}
\end{equation}
The expression makes the signal--background tradeoff explicit: a narrow window
preserves a larger fraction of detections from a low-jitter receiver while
admitting less stationary background.  A measured asymmetric
tail must be integrated directly rather than replaced by
Eq.~\eqref{eq:timing_capture}.

For synchronization and ranging, system timing includes more than the detector.
Clock recovery, time-transfer calibration, propagation variation, centroid bias,
and FPGA/TDC quantization can exceed the intrinsic jitter
\cite{Pachedjieva2018SpaceAccuracy,Trojanek2018TimeTransfer,Srinivasan2018Centroiding}.  Real-time photon-counting receivers consequently
characterize the complete time-tagging chain, including detector, comparator,
TDC, synchronization markers, and FPGA processing
\cite{Vyhnalek2019Realtime,Caplan2024PhotonCounting,Marsden2024GroundReceiver}.
For an array, per-pixel timing offsets and differential nonlinearity of the TDC
must be calibrated; reporting only the best pixel hides the error seen by an
imaging or tracking algorithm \cite{Chen2024SPADReceiver,Licht2025PhotonCountingCamera,Bashir2021BeamTracking}.

Dead time begins when an event prevents the same detection channel from resolving
another event.  The measurement must specify whether the quoted interval is an
electrical output pulse width, enforced hold-off, time to a chosen recovery
fraction, or the minimum resolvable interarrival time.  A SPAD may be fully blind
during active hold-off and then return abruptly, whereas a passively recharging
SPAD, SiPM cell, or SNSPD can recover sensitivity continuously.  Interarrival-time
histograms and a two-pulse experiment reveal this distinction more reliably than
the reciprocal of a peak count rate.  Count-rate measurements should also state
whether the input was uniform, focused on one pixel, or distributed over the
 array, because pixel dead time, shared arbitration, TDC bandwidth, and output I/O
 may impose different limits \cite{Sarbazi2018SPADStatistics,Huang2023SPADOFDM,Huang2022FiveGbpsSPAD}.  Underwater and time-controlled
 receivers provide additional recovery regimes
 \cite{Lyu2022SPADUWOC,Liu2023TimeControlledSPAD,Liu2025NovelCMOSSPAD}.  High-dynamic-range
receivers exploit multiple pixels or channels precisely because aggregate rate
cannot be inferred from a single-pixel reset time
\cite{Rogalin2021HighDynamicRange,Hao2024SNSPDArray,Du2018MultiChannel,Lai2025InGaAsSPAD}.

\subsubsection{Saturation, afterpulsing, crosstalk, and readout loss}

Linearity should be characterized by sweeping calibrated photon flux over the
entire intended operating range and plotting both the measured rate and the ratio
of measured to ideal rate.  The onset criterion---for example, 1\% compression,
10\% compression, or a BER penalty---must be stated.  For a single SPAD, binary
occupancy and dead time cause compression; for a SiPM, photons compete for a
finite number of charged microcells; and for an SNSPD, incomplete current recovery
 can lower detection probability before pulses cease to be individually resolved.
 The communication demonstrations in visible, NIR, and underwater links show that
 useful rate depends on modulation and signal processing as well as on detector
 hardware \cite{Huang2022FiveGbpsSPAD,Huang2023SPADOFDM,Ali2023SiPM82Gbps}.  Equalized and multi-mode SiPM/SPAD links reinforce that
 dependence \cite{Zhang2023SiPMEqualization,Li2025SiPMModes,Lyu2022SPADUWOC}.
 A saturation curve should
therefore accompany the low-flux PDE value.

A useful first-order model is the nonparalyzable approximation
\begin{equation}
  \frac{R_{\mathrm{out}}}{R_0}=
  \left(1+\frac{R_0\tau}{N_{\mathrm{eff}}}\right)^{-1},
  \label{eq:parallel_deadtime}
\end{equation}
where $R_0$ is the pre-dead-time event rate, $\tau$ is the recovery scale, and
$N_{\mathrm{eff}}$ is the number of independently illuminated channels.  The
curves use 3~ns for a single silicon SPAD, 3.5~ns and 16 pixels for a recent CMOS
 receiver, 50~ns and 64 channels for the InGaAs receiver demonstration, 45~ns and
 5676 microcells for the SiPM, and 1~ns and four pixels for the compact SNSPD
 \cite{Gramuglia2021LowNoiseSPAD,Liu2025NovelCMOSSPAD,Lai2025InGaAsSPAD}.
 The SiPM and SNSPD parameters follow representative communication-oriented
 studies \cite{Zhang2022SiPMGuideline,Hao2024SNSPDArray}.  Equation~\eqref{eq:parallel_deadtime}
assumes uniform loading and excludes arbitration and output-bandwidth loss; a
focused spot reduces $N_{\mathrm{eff}}$ and moves compression to a lower rate.

Afterpulsing is conditional noise: carriers trapped during a primary avalanche
are released later and may trigger a secondary event.  A practical estimate uses
an interarrival-time histogram, subtracts the independently measured accidental
background, and integrates the excess probability over a declared time window.
Its value depends strongly on hold-off, excess bias, temperature, and count rate.
Optical crosstalk occurs when avalanche-generated photons trigger neighboring
microcells; electrical crosstalk and common readout paths can produce similar
coincidences.  Consequently, a quoted crosstalk probability must specify whether
it was obtained from pulse-amplitude statistics, pixel coincidences, or an
application-level fit \cite{Chitnis2014SPADSystem,Fisher2017CMOSSPAD,Wang2019SPADOperation,He2022SiPMSelection}.  For SiPMs, afterpulsing and crosstalk
alter both mean charge and excess noise, so photon-number peaks cannot be treated
as independent photons without correction \cite{Essalih2020SiPMUWOC,Zhang2020SiPMUWOC}.

Digital arrays introduce a distinct loss mechanism after successful optical
detection.  Simultaneous pixel events may collide in an OR tree or arbiter, share
a finite counter depth, or exceed TDC and output-link bandwidth.  This
\emph{readout loss} must not be reported as reduced PDE unless the measurement
reference plane is explicitly the receiver output.  Uniform flood illumination,
a moving focused spot, and sparse pulsed illumination stress different resources.
 Integrated receivers and imaging arrays demonstrate why per-pixel dead time,
 column contention, timestamp throughput, and aggregate output rate should be
 measured separately \cite{Fisher2013IntegratingReceiver,Li2022IntegratedSPAD,Chen2024SPADReceiver}.  Camera and diffraction-limited detector studies add
  focal-plane evidence \cite{Licht2025PhotonCountingCamera,Lit0240}.  Statistical and
machine-learning receivers can compensate some deterministic compression, but
they do not recreate events that were never recorded
\cite{Sarbazi2018SPADStatistics,Arya2022MLBoundaries,Peppas2017UWOCSpatialDiversity}.

\subsubsection{Array, packaging, and system-level metrics}

The minimum characterization record must report the operating point beside each metric.
For arrays this includes pixel count, active area, pitch, fill factor, enabled
pixel map, nonuniformity, hot-pixel rejection, per-pixel or shared TDC structure,
and sustainable output throughput.  Imaging and tracking receivers additionally
 need point-spread function, centroid accuracy, maximum spot velocity, FOV, and the
 latency from photon arrival to tracking command
  \cite{Srinivasan2018Centroiding,Bashir2021BeamTracking,Chen2024SPADReceiver}.  Imaging-device studies further motivate reporting the
  active focal-plane configuration \cite{Licht2025PhotonCountingCamera,Lit0240}.
  A high aggregate count rate
measured under uniform illumination does not guarantee the same rate for a spot
concentrated on a few pixels.

Packaging closes the gap between a laboratory device and an OWC receiver.  For a
fiber input, insertion loss, connector repeatability, supported mode count, and
polarization sensitivity must be recorded; for free-space coupling, the relevant
quantities include active diameter, acceptance angle, focus tolerance, alignment
drift, and obscuration.  Few-mode coupling and photonic-lantern experiments show
that increased collected étendue can improve robustness while adding insertion
loss and multiple detector/readout channels
\cite{Vyhnalek2019FewMode,Tedder2020PhotonicLantern}.  Narrowband filters and
adaptive optics similarly exchange throughput and complexity for background or
coupling improvement \cite{Shtyrkova2019Narrowband,Mikos2026CoherentCombination}.

Cooling must be reported at the system level.  ``Operating at 4~K'' is not a SWaP
metric unless cryocooler input power, heat lift, cooldown time, vibration,
orientation, control electronics, mass, and volume are included.  This distinction
is essential when comparing a passively cooled SPAD/SiPM payload with an SNSPD
ground receiver \cite{Grein2015Moon,Wu2017DeepSpace,Biswas2024DeepSpacePhoton}.  Space-to-ground receiver programs further show that
maintainability, detector modularity, timing interfaces, and real-time processing
can determine operational value even when detector sensitivity is excellent
\cite{Vyhnalek2018ModularSNSPD,Grein2015LLCDReceiver,Downey2024GroundReceiver,Marsden2024GroundReceiver}.

Finally, the metric set must match the intended link.  Indoor and underwater
terminals often value large area, passive cooling, and background tolerance
\cite{Mapunda2020IndoorVLC,Hu2018UWOC256PPM,Chen2020LongDistanceUWOC}; deep-space
receivers prioritize SDE, DCR, timing, and coded photon efficiency
\cite{Caplan2006PhotonEfficient,Dolinar2012UltimateLimits,Boroson2018PerformanceLimits}; NLOS and joint communication--tracking systems
place additional weight on FOV, array uniformity, and spatial observability
\cite{Gong2015NLOS,Wang2023NLOSPositioning,Sun2025AirWater}.  Thus, a fair
technology comparison is not a ranking by one column.  It is a vector of
co-measured quantities tied to the same optical reference plane, operating point,
readout configuration, and communication task.

\subsection{Communication-Oriented Nonideal Models and Engineering Co-Design}
\label{subsec:detector_models_selection}

Detector modeling and detector selection should be one continuous calculation.
A link budget supplies a time-, space-, and wavelength-dependent photon flux;
the detector and readout transform that flux into a random digital record; and
the modem converts the record into synchronization statistics, likelihoods, and
decoded bits.  Separating these steps is useful analytically, but selecting a
device from a stand-alone PDE or DCR value breaks the chain.  This subsection
therefore develops a layered model from incident photons to recorded counts and
then uses the same parameters as engineering selection constraints.  The aim is
not one universal closed form.  It is a model hierarchy in which complexity is
added only when a nonideality changes the design conclusion.

\subsubsection{Ideal reference and the unified event chain}

Consider pixel or detection channel $i$ over an observation interval
$\mathcal T$.  Conditioned on transmitted symbol $x$, the ideal primary-event
mean is
\begin{equation}
  \Lambda_i(x)=\int_{\mathcal T}
  \left[\eta_i(t,\lambda)s_i(t,\lambda|x)
  +b_i(t,\lambda)+d_i(t)\right],\mathrm dt,
  \label{eq:channel_primary_mean}
\end{equation}
where $s_i$, $b_i$, and $d_i$ denote incident signal, registered background,
and dark-event rates, respectively.  For independent arrivals and a fully
recovered detector,
\begin{equation}
  K_i|x\sim\operatorname{Poisson}\!\left(\Lambda_i(x)\right).
  \label{eq:ideal_poisson_channel}
\end{equation}
For a binary gated SPAD, Eq.~\eqref{eq:ideal_click_probability} follows by
mapping all $K_i\geq1$ outcomes to a single click.  For a counting receiver,
$K_i$ is retained; for an ideal time-tagging receiver, the arrival times inside
$\mathcal T$ are retained as well.  Equations~\eqref{eq:channel_primary_mean}
 and \eqref{eq:ideal_poisson_channel} are the correct baseline for photon-starved
 links, but only after signal, background, and dark rates have been referenced to
 the same optical plane \cite{Karp1970PhotonCounting,Yiannopoulos2019ErrorProbability,Sarbazi2018SPADStatistics}.
 Classical $M$-ary Poisson detection, threshold mean-photon-number modulation,
 and shot-noise-aware amplitude design provide complementary reference cases
 \cite{Lit0057,Lit0171,Lit0233}.

A practical receiver can be represented as a cascade
\begin{equation}
  \mathbf Y=
  \mathcal Q_{\mathrm{ro}}
  \!\left\{\mathcal C_{\mathrm{xt}}
  \!\left[\mathcal A_{\mathrm{ap}}
  \!\left(\mathcal R_{\tau,\mathbf N}(\mathbf K)\right)\right]\right\},
  \label{eq:unified_detector_chain}
\end{equation}
where $\mathcal R_{\tau,\mathbf N}$ represents dead time, recovery, and finite
available cells; $\mathcal A_{\mathrm{ap}}$ adds history-dependent afterpulses;
$\mathcal C_{\mathrm{xt}}$ couples events between cells or channels; and
$\mathcal Q_{\mathrm{ro}}$ represents discrimination, time quantization,
arbitration, buffering, and output-link loss.  The ordering is physically
important.  A photon rejected by coupling or PDE never becomes $K_i$; an event
lost in an arbiter occurred in the device but never reaches $Y_i$; and neither
loss can be recovered exactly by DSP.  Equation~\eqref{eq:unified_detector_chain}
also prevents unlike effects from being hidden inside an empirical ``effective
PDE.''

\subsubsection{Dead time, occupancy, and recovery}

For a stationary incident primary-event rate $R_0$, two standard single-channel
limits are
\begin{equation}
  R_{\mathrm{np}}=\frac{R_0}{1+R_0\tau_{\mathrm d}},
  \qquad
  R_{\mathrm p}=R_0\exp(-R_0\tau_{\mathrm d}),
  \label{eq:deadtime_models}
\end{equation}
for nonparalyzable and paralyzable dead time, respectively.  The first model is
a useful approximation to a fixed hold-off; the second describes a detector for
which arrivals during the blind interval extend the blindness.  Neither is
automatically valid for a passively recharging SPAD or SNSPD, because detection
efficiency and pulse amplitude may recover continuously.  In that case the
state variable should include the time since the last event, with a measured
recovery function $g(\Delta t)$ multiplying the instantaneous PDE
\cite{Sarbazi2018SPADStatistics,Huang2021TimeGated,Liu2023TimeControlledSPAD,Det0091}.

For an array, Eq.~\eqref{eq:parallel_deadtime} replaces one recovery channel by
an effective number $N_{\mathrm{eff}}$ of independently illuminated channels.
If no more than a fraction $\epsilon$ of events may be lost under the
nonparalyzable approximation, a first sizing rule is
\begin{equation}
  N_{\mathrm{eff}}\geq
  \frac{1-\epsilon}{\epsilon}R_0\tau_{\mathrm d}.
  \label{eq:array_sizing_rule}
\end{equation}
This is a lower bound rather than a pixel-count requirement: a focused spot,
dead or disabled pixels, microlenses, and nonuniform PDE can make
$N_{\mathrm{eff}}$ much smaller than the physical array size.  For SiPMs, the
occupancy law in Eq.~\eqref{eq:sipm_occupancy} must additionally be convolved
with microcell recharge when the waveform duration is comparable to the recovery
time.  Thus a short pulse is primarily limited by the number of charged cells,
whereas a long waveform also depends on how often cells can refire
\cite{Essalih2020SiPMUWOC,Zhang2022SiPMGuideline,Zhang2023SiPMEqualization,Li2025SiPMModes}.

The communication consequence is channel memory.  If $\tau_{\mathrm d}$ or the
recharge tail overlaps the next symbol, the conditional distribution is
$p(y_k|x_k,z_k)$, where $z_k$ is the detector state inherited from previous
symbols, rather than the memoryless $p(y_k|x_k)$.  Reducing the symbol interval
can therefore lower rather than raise useful throughput.  Interleaving pixels,
 time gating, optical attenuation, waveform predistortion, and likelihoods that
 track $z_k$ address different parts of this problem
 \cite{Huang2023SPADOFDM,Det0016,Det0095}.  Adaptive attenuation and sub-dead-time
 operation address the remaining overload regime \cite{Det0123,Det0167}.

\subsubsection{Array nonuniformity and readout loss}

For an $M$-pixel detector, the vector $\boldsymbol\Lambda(x)$ should retain the
spatial illumination profile and per-pixel efficiency, background, DCR, dead
time, and timing offset.  Replacing it by its sum is safe only when pixels are
exchangeable and the readout preserves all simultaneous events.  In practice,
edge pixels, hot-pixel masking, microlens response, spot motion, and temperature
gradients produce a nonuniform array.  Calibration should therefore estimate a
parameter set $\{\eta_i,d_i,\tau_i,\Delta t_i\}_{i=1}^{M}$ and propagate the
 actual point-spread function across it, especially for imaging and tracking
 receivers \cite{Chen2024SPADReceiver,Licht2025PhotonCountingCamera,Lit0240,Bashir2021BeamTracking}.  Beacon-free alignment and joint optical-surface
 precoding demonstrate how spatial detector state couples back into transmitter
 control \cite{Lit0038,Lit0174}.

Readout loss is modeled after device-level detection.  A useful diagnostic is
\begin{equation}
  \eta_{\mathrm{ro}}(R,\boldsymbol\rho)=
  \frac{\mathbb E[N_{\mathrm{reported}}]}
       {\mathbb E[N_{\mathrm{device}}]},
  \label{eq:readout_efficiency}
\end{equation}
where $R$ is the total device-event rate and $\boldsymbol\rho$ describes its
distribution across pixels, columns, TDCs, and time.  The dependence on
$\boldsymbol\rho$ distinguishes uniform flood illumination from a tracked spot:
both can produce the same total event rate but very different arbiter collision
and column-load probabilities.  Measured or simulated service curves are more
 reliable than a single maximum-throughput number when events share an OR tree,
 TDC bank, buffer, serializer, or FPGA input
 \cite{Fisher2013IntegratingReceiver,Li2022IntegratedSPAD,Det0385}.
 Integrated and time-coordinated arrays expose related bottlenecks
 \cite{Det0390,Det0164}.  The interface should be enlarged, load-balanced, or
locally aggregated whenever $\eta_{\mathrm{ro}}$, rather than the detector PDE,
sets the receiver sensitivity.

\subsubsection{Background and correlated false events}

Signal, background, and DCR are independent only at the primary-event level.
Afterpulsing introduces temporal correlation with a previous avalanche, while
optical or electrical crosstalk introduces correlation between cells.  A compact
conditional-intensity model for channel $i$ is
\begin{equation}
\begin{split}
  \lambda_i(t|\mathcal H_t)=&\;g_i(t-t_i^{\mathrm{last}})
  [\lambda_{s,i}(t)+\lambda_{b,i}(t)+d_i]\\
  &+\sum_{t_{i,k}<t}h_{\mathrm{ap},i}(t-t_{i,k})
  +\sum_{j\ne i}c_{ji}\sum_k\delta(t-t_{j,k}),
\end{split}
\label{eq:correlated_event_intensity}
\end{equation}
where $\mathcal H_t$ is the event history, $g_i$ is recovery, $h_{\mathrm{ap},i}$
is an afterpulse kernel, and $c_{ji}$ is an effective crosstalk coupling.  This
model can be simplified to independent Poisson counts only when recovery is
complete and the correlated terms are negligible.  Otherwise, fitting a Poisson
mean can reproduce the average count while underestimating variance and burst
errors \cite{Chitnis2014SPADSystem,Wang2019SPADOperation,He2022SiPMSelection,Sarbazi2018SPADStatistics}.

Background mitigation acts before or during detection: narrowband filters reduce
spectral radiance, small FOV or single-mode coupling reduces admitted spatial
modes, time gating reduces the observation interval, and pixel selection rejects
regions outside the tracked point-spread function.  Hold-off suppresses some
afterpulses but also sacrifices live time; a higher discriminator threshold can
reduce false pulses but may reject partially recovered events.  These tradeoffs
must be evaluated using the installed aperture and optical train
\cite{Shtyrkova2019Narrowband,Rogalin2021HighDynamicRange,Wan2024DaytimeSNSPD,Det0134}.  Subtracting an average background in DSP cannot
restore saturation headroom or timestamps already lost to background-triggered
dead time.

\subsubsection{Consequences for BER, information rate, and synchronization}

For equiprobable binary signaling and decision region $\mathcal D_1$, the BER is
\begin{equation}
  P_{\mathrm e}=\frac{1}{2}
  \Pr(Y\in\mathcal D_1|X=0)+\frac{1}{2}
  \Pr(Y\notin\mathcal D_1|X=1).
  \label{eq:ber_general_detector}
\end{equation}
An ideal Poisson threshold detector evaluates these probabilities from two
Poisson distributions.  A nonideal receiver must instead obtain them from the
output $\mathbf Y$ of Eq.~\eqref{eq:unified_detector_chain}, including detector
state when dead time spans symbols.  Saturation can compress the separation of
the two distributions; afterpulsing and crosstalk can broaden them; background
raises both means; and readout loss can make the variance depend on the spatial
profile.  Consequently, a threshold optimized under a Poisson assumption may be
biased even when its predicted average count is correct
\cite{Yiannopoulos2019ErrorProbability,Det0117,Det0388,Huang2023SPADOFDM}.

The same issue appears in information-rate claims.  In the memoryless limit, one
may optimize $I(X;Y)$ using a Poisson or Bernoulli channel.  With recovery state
$Z_k$, the relevant quantity is a finite-state channel information rate,
\begin{equation}
  R_{\mathrm I}=\liminf_{n\rightarrow\infty}
  \frac{1}{n}\max_{p(x^n)} I(X^n;Y^n),
  \label{eq:finite_state_information_rate}
\end{equation}
subject to optical power, bandwidth, and peak constraints.  Treating correlated
or saturated counts as independent can overstate both capacity and coding gain.
Photon-efficient PPM and coded deep-space links operate near the opposite, low-
occupancy limit, where background, erasures, and synchronization overhead often
 matter more than dead-time compression
 \cite{Caplan2006PhotonEfficient,Dolinar2012UltimateLimits,Banaszek2020QuantumLimits,Boroson2018PerformanceLimits}.
 Rate and power-allocation studies for underwater and low-photon-rate channels
 show that the operating optimum can move substantially with background and
 user loading \cite{Lit0169,Lit0190,Lit0196}.  Model-aware end-to-end learning
 and iterative Poisson-channel estimation offer data-driven alternatives when
 an analytic likelihood is incomplete \cite{Lit0251,Lit0259}.

Synchronization depends on the shape of the event-time distribution, not only
its total count.  Detector jitter convolves the received waveform, background
adds nearly uninformative time tags, gating can truncate the timing peak, and
dead time can distort a strong acquisition preamble.  Clock and frame estimators
should therefore be tested with the same flux, background, gate, and detector
state expected during operation \cite{Det0006,Det0014,Det0156,Vyhnalek2019Realtime}.  A design that meets uncoded BER after perfect alignment
may still fail because it cannot acquire or track the link quickly enough.

\subsubsection{Selection and integration workflow}

A practical selection process begins by declaring the receiver optical
port and scenario envelope: wavelength, aperture, FOV, spot distribution,
background radiance, data rate, modulation, acquisition time, pointing error,
and environmental limits.  The link budget is then converted into distributions
of signal and background photon rate, not only nominal means.  Candidate devices
are retained only if their calibrated spectral efficiency, timing, active area,
recovery, array architecture, and temperature range cover that envelope.

For every surviving candidate, the measured parameters should be propagated
through Eqs.~\eqref{eq:unified_detector_chain}--\eqref{eq:ber_general_detector}
to obtain BER, acquisition probability, sustainable throughput, and margin over
the complete operating envelope.  The array and readout are then sized for a
high percentile of received flux using Eq.~\eqref{eq:array_sizing_rule}, not only
the average link budget.  Optical coupling, filter drift, calibration, FPGA/TDC
capacity, cooling, power, mass, and maintainability are added before ranking.
This feasibility-first procedure prevents a high-PDE detector from winning a
paper comparison while failing the intended terminal.

The resulting choices are scenario dependent.  CMOS SPAD arrays are strong when
time gating, spatial filtering, tracking, and compact digital integration are
valuable; SiPMs are attractive when large area and simple analog summation are
more important than per-pixel information; and SNSPDs are compelling when low
background, telecom-band efficiency, and timing justify cryogenic infrastructure.
PMT/MCP-PMT receivers remain relevant for large-area or scattering geometries,
while upconversion can be useful when infrared wavelength and controlled spatial
modes permit an efficient nonlinear front end.  Hybrid paths are preferable when
the flux envelope is too wide for one detector mode
\cite{Huang2023SixG,Caplan2024PhotonCounting,Det0062,Guardiani2024SNSPDLaserCom}.

The final comparison should therefore report a \emph{port-to-bit} result: optical
power and background at a declared receiver port, the detector/readout operating
point, synchronization success, coded BER or frame-error rate, sustained data
rate, and the associated SWaP and thermal burden.  This common endpoint unifies
the nonideal models and engineering guidelines.  It also makes clear when better
performance comes from the detector itself and when it comes from filtering,
coupling, parallelism, readout, or DSP.
\section{Photon-Counting Modulation and Receiver Signal Processing}
\label{sec:modulation_signal_processing}

\subsection{Photon-Counting Channel Models and Modulation Formats}
\label{subsec:channel_modulation}

Once photons have been converted into registered events, the communication
problem is to choose a waveform and an observation that preserve the information
carried by those events.  Let symbol $X=m$ select a nonnegative received-power
waveform $P_m(t)$ over an interval $0\leq t<T_{\mathrm s}$.  At the output of an
ideal linear detector, the corresponding registered-event intensity can be
written
\begin{equation}
  \lambda_m(t)=\frac{\eta_{\mathrm{sys}}P_m(t)}{h\nu}
  +\lambda_{\mathrm b}(t)+\lambda_{\mathrm d},
  \label{eq:modulation_event_intensity}
\end{equation}
where $\eta_{\mathrm{sys}}$ is referenced to a declared receiver port,
$\lambda_{\mathrm b}(t)$ is the admitted background rate, and
$\lambda_{\mathrm d}$ is the dark-event rate.  If the receiver retains only the
total count $Y$ and the rate is conditionally Poisson, then
\begin{equation}
  \Pr(Y=y\mid X=m)=\frac{\mu_m^y e^{-\mu_m}}{y!},\qquad
  \mu_m=\int_0^{T_{\mathrm s}}\lambda_m(t)\,\mathrm dt .
  \label{eq:poisson_modulation_channel}
\end{equation}
This model is the natural reference for photon-counting optical links and is
accurate when arrivals are independent, the occupancy per resolving element is
small, and the detector has recovered before the next relevant event
\cite{Karp1970PhotonCounting,Yiannopoulos2019ErrorProbability,Boroson2018PerformanceLimits}.  It should not be treated as a universal detector
law.  Dead time, recharge, correlated events, and finite readout bandwidth turn
Eq.~\eqref{eq:poisson_modulation_channel} into the state-dependent channel
developed in Section~\ref{sec:detector_technologies}; the Poisson model is then a
benchmark against which the implementation loss is measured.

A time-tagging receiver retains more information than a symbol counter.  For
registered event times $\mathcal T=\{t_1,\ldots,t_N\}$, the log-likelihood of
symbol $m$, apart from a term independent of $m$, is
\begin{equation}
  \ln p(\mathcal T\mid m)=
  \sum_{i=1}^{N}\ln \lambda_m(t_i)
  -\int_0^{T_{\mathrm s}}\lambda_m(t)\,\mathrm dt .
  \label{eq:timestamp_log_likelihood}
\end{equation}
The total count is sufficient when the candidate waveforms produce the same
normalized temporal shape and differ only in integrated intensity.  It is not
sufficient when information is carried by pulse position, when the gate clips a
pulse, or when synchronization is uncertain.  In those cases, slot counts or
individual timestamps preserve the timing information needed for joint
acquisition and demodulation.  This distinction is particularly important in
photon-efficient PPM receivers and in links that use the same photon stream for
communication and tracking
\cite{Caplan2006PhotonEfficient,Det0010,Det0156,Licht2025PhotonCountingCamera}.

The observation alphabet also depends on detector architecture.  For $J$
uniformly illuminated binary pixels that can each fire at most once during a
gate, the ideal Poisson count is replaced by
\begin{equation}
  Y\mid X=m\sim\operatorname{Binomial}
  \left(J,1-e^{-\mu_m/J}\right).
  \label{eq:finite_pixel_modulation_channel}
\end{equation}
The $J=1$ case is a click/no-click channel, whereas
Eq.~\eqref{eq:finite_pixel_modulation_channel} approaches the Poisson law as
$J\rightarrow\infty$ at fixed $\mu_m$.  A nonuniform spot produces a
Poisson-binomial channel, and a recovering array requires a state for the number
and spatial distribution of available pixels.  Thus, a modulation format that
is well separated in incident optical power may become poorly separated after
the nonlinear mapping to reported counts.  Experimental and
statistical studies of SPAD and SiPM links likewise show that the receiver model
must follow the detector operating regime rather than defaulting to additive
Gaussian noise \cite{Sarbazi2018SPADStatistics,Det0092,Det0393}.

\subsubsection{Binary and multilevel intensity modulation}

On--off keying (OOK) assigns two mean counts,
$\mu_0=\mu_{\mathrm b}$ and $\mu_1=\mu_{\mathrm b}+\mu_{\mathrm s}$, to the
off and on symbols.  For equal priors and $\mu_1>\mu_0>0$, an ideal Poisson
receiver decides ``on'' when
\begin{equation}
  y\ln\!\left(\frac{\mu_1}{\mu_0}\right)-(\mu_1-\mu_0)>0,
  \label{eq:ook_poisson_test}
\end{equation}
with an integer threshold obtained from the left-hand side.  OOK is attractive
because its modulator, synchronization, and counter can be simple, and it does
not require the large peak-to-average ratio of high-order PPM.  Its threshold,
however, depends directly on background and gain calibration.  A click-only
receiver also loses amplitude separation when both $1-e^{-\mu_0}$ and
$1-e^{-\mu_1}$ approach unity, while dead time creates data-dependent memory at
high symbol rates.

$M$-ary pulse-amplitude modulation (PAM) increases the number of bits per symbol
by assigning several nonnegative intensity levels.  Under the ideal model, its
adjacent decision boundaries follow from pairwise versions of
Eq.~\eqref{eq:ook_poisson_test}; with nonuniform priors, the log-prior ratio is
added to the test.  PAM uses time efficiently and is consequently attractive in
bandwidth-limited, moderate-to-high-count links.  The price is reduced spacing
between conditional count distributions and strong sensitivity to saturation,
power drift, and background-estimation error.  Learned decision boundaries can
track a nonlinear multilevel receiver, but training does not remove the
 information loss that occurs after two optical levels have been mapped to nearly
 the same count distribution \cite{Arya2022MLBoundaries}.  Array size, optical
 spot uniformity, and automatic attenuation should therefore be co-designed with
 the PAM order \cite{Rogalin2021HighDynamicRange,Det0123}.
 Neural demodulators and greedy photon-efficient receivers provide alternative
 decision rules, but both still depend on a calibrated count distribution
 \cite{Lit0140,Lit0237}.

Multicarrier waveforms occupy a related high-throughput regime.  OFDM can use the
electrical bandwidth efficiently and can simplify equalization of dispersive
optical channels, but its large instantaneous peaks interact with SPAD blocking
and SiPM saturation.  A Gaussian-noise approximation may be adequate after many
independent pixels contribute moderate counts; it is unreliable in the sparse
or strongly clipped regime.  Demonstrations of high-rate SPAD links therefore
 combine array parallelism, carefully chosen operating points, and receiver-side
 nonlinearity management rather than relying on the waveform alone
 \cite{Huang2023SPADOFDM,Huang2022FiveGbpsSPAD}.
 OFDM clipping analysis, laboratory comparisons with single-carrier/PAM
 signaling, and adaptive bias control quantify the transmitter--detector
 interaction \cite{Lit0235,Lit0246,Lit0247}.

\subsubsection{Pulse-position modulation and photon efficiency}

In $M$-ary PPM, one energetic pulse is placed in one of $M$ slots, conveying
$\log_2 M$ uncoded bits per symbol.  With equal background in all slots and
known timing, the ideal maximum-likelihood receiver selects the slot with the
largest count, with an explicit rule for ties.  Concentrating energy into one
slot separates the signal observation from empty slots and can provide high
photon information efficiency (PIE), which explains the use of coded PPM in
deep-space, long-range, and strongly attenuating underwater links
\cite{Grein2015LLCDReceiver,Hu2018UWOC256PPM,Biswas2024DeepSpacePhoton,Lai2025InGaAsSPAD}.  Upconversion receivers have also
been demonstrated with PPM at telecom wavelengths, showing that modulation and
front-end technology can be selected jointly \cite{Det0049}.

The gain is not free.  If slot duration is fixed, the PPM symbol occupies $M$
slots and its raw bit rate scales only as $\log_2(M)/M$ per slot.  If symbol
duration is fixed instead, increasing $M$ demands proportionally finer timing
resolution and electrical bandwidth.  In either interpretation, the peak-to-
average optical power grows, synchronization becomes more demanding, and every
additional empty slot admits background.  Dead time may erase a signal following
a background event, while afterpulsing can transfer evidence into later slots.
Guard intervals reduce this memory but consume rate.  Multi-pulse PPM (MPPM)
 places $w$ pulses among $M$ slots and carries up to
 $\log_2\binom{M}{w}$ bits per symbol, providing an intermediate tradeoff among
 peak power, spectral efficiency, and detection complexity
 \cite{Peppas2017UWOCSpatialDiversity}.
 Efficient arbitrary-order PPM error evaluation is useful when choosing this
 order under a measured background level \cite{Lit0175}.
 Polarized PPM, multi-header PPM, and differential PPM illustrate three ways to
 trade peak power, dimming support, and receiver memory
 \cite{Lit0052,Lit0191,Lit0238}.

\begin{figure}[!t]
  \centering
  \includegraphics[width=\columnwidth]{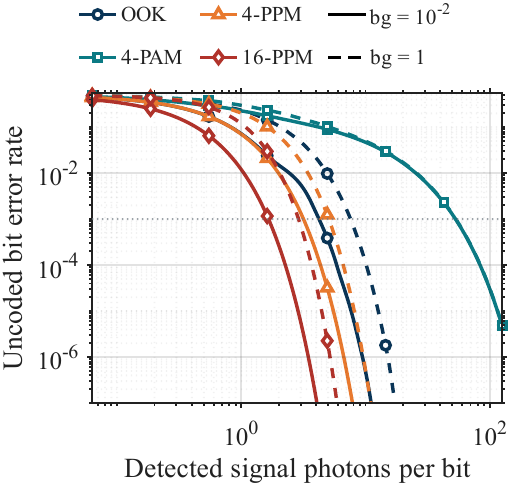}
  \caption{Uncoded BER versus registered signal photons per information bit for
  ideal Poisson OOK, 4-PAM, 4-PPM, and 16-PPM reception. OOK is evaluated by
  importance-sampled Monte Carlo with $10^6$ trials per photon-number anchor;
  the other curves use exact Poisson probability sums. Solid and dashed curves
  denote $10^{-2}$ and 1 registered background photons per slot, respectively;
  the gray guide marks BER $=10^{-3}$.}
  \label{fig:modulation_required_photons}
\end{figure}

Figure~\ref{fig:modulation_required_photons} compares the complete uncoded BER
curves at $10^{-2}$ and 1 registered background photons per slot.  The OOK
curves use importance-sampled Monte Carlo with $10^6$ trials per photon-number
anchor and the exact scalar-count MAP rule.  This rare-event construction
maintains statistical resolution well below BER $=10^{-5}$, where direct Monte
Carlo would observe too few errors.  Gray-labeled 4-PAM uses exact bitwise MAP
probabilities, whereas PPM uses the maximum-count decision with random resolution
of ties.  Every PPM symbol contains one signal pulse and $M$ background-bearing
slots.  The result isolates modulation behavior under ideal counting, before dead time, saturation,
and synchronization overhead are added.  Four-PAM now exhibits its gradual BER
decrease as adjacent count levels separate, while the two line styles show the
rightward sensitivity penalty caused by background.  The gray guide permits the
required photons per bit to be read at BER $=10^{-3}$.

No modulation order is therefore optimal independently of the resource
constraint.  Two useful normalized measures are
\begin{equation}
  \mathrm{PIE}=\frac{I(X;Y)}{\mathbb E[N_{\mathrm s}]},
  \qquad
  \mathrm{SE}=\frac{I(X;Y)}{B T_{\mathrm s}},
  \label{eq:pie_se_definitions}
\end{equation}
where $N_{\mathrm s}$ is the registered signal count and $B$ is the occupied
receiver bandwidth.  High-order PPM approaches the photon-efficient, low-
spectral-efficiency corner; PAM and multicarrier signaling move toward higher
spectral efficiency but demand more distinguishable counts per interval.  This
 tradeoff is consistent with the fundamental limits of photon-counting optical
 communication \cite{Dolinar2012UltimateLimits,Banaszek2020QuantumLimits} and
 with capacity analyses for general quantum Gaussian receivers \cite{Lit0205}.

A fair numerical comparison must hold fixed the declared receiver port, average
and peak optical power, net information rate, bandwidth, background radiance and
gate width, and detector/readout model.  It must also include empty slots, guard
times, synchronization overhead, and coding rate.  Comparing OOK and PPM only at
equal photons per symbol can otherwise favor the format that transmits fewer net
bits or occupies more time.  In practice, OOK is a robust baseline, PAM and OFDM
serve bandwidth-limited high-count links, and coded PPM serves the photon-starved
regime; the boundaries move when background, saturation, and timing constraints
are applied.  The receiver algorithms in the next subsection should consequently
operate on the likelihood associated with the selected detector and modulation,
not on a waveform-independent noise assumption.

\subsection{Photon-Counting Detection, Synchronization, and Channel Estimation}
\label{subsec:detection_sync_estimation}

Detection, synchronization, and parameter estimation are often presented as
separate receiver blocks, but they act on the same sparse event record.  A
decision threshold requires the background and signal gain; a channel estimator
requires symbol or pilot timing; and a timing metric depends on the assumed
detector likelihood.  This coupling is weak only in a high-count, nearly linear
receiver for which a Gaussian approximation is accurate.  It is pronounced in
photon-starved and occupancy-limited links, including SPAD, SiPM, PMT, and SNSPD
implementations \cite{Karp1970PhotonCounting,Yiannopoulos2019ErrorProbability,Boroson2018PerformanceLimits,Huang2023SixG}.  The appropriate interface between the optical front
end and DSP is therefore a calibrated conditional probability law rather than a
single nominal responsivity.

Let $\mathbf y=(y_1,\ldots,y_K)$ denote gated counts or binned timestamps over a
candidate observation interval.  For pilot or data samples $x_k$, channel gain
$g$, admitted background $b$, timing offset $\tau$, and detector state
$s_k$, define
\begin{equation}
\begin{aligned}
  \mu_k(g,b,\tau)&=b+g\,x_k(\tau),\\
  \ell(g,b,\tau)&=\sum_{k=1}^{K}
  \ln p_{\mathcal D}(y_k\mid \mu_k,s_k),
\end{aligned}
  \label{eq:joint_receiver_likelihood}
\end{equation}
where $p_{\mathcal D}$ is the response law of detector $\mathcal D$.  It is
Poisson in the ideal low-occupancy limit, binomial or Poisson-binomial for a
one-click-per-pixel gate, and state dependent when dead time, recharge, event
correlation, or arbitration couples samples.  Equation
\eqref{eq:joint_receiver_likelihood} provides one framework for symbol MAP
detection, generalized likelihood ratio tests (GLRTs), joint estimation, and
soft information generation.  Statistical SPAD models, integrating receivers,
and practical high-dynamic-range architectures all support this detector-aware
view \cite{Sarbazi2018SPADStatistics,Fisher2017CMOSSPAD,Li2022IntegratedSPAD,Det0385}.

\subsubsection{Count-domain detection and soft outputs}

With known $g$, $b$, and $\tau$, symbol-by-symbol MAP detection selects
\begin{equation}
  \widehat m=\arg\max_m
  \left\{\ln p_{\mathcal D}(\mathbf y\mid X=m)
  +\ln P_X(m)\right\}.
  \label{eq:detector_aware_map}
\end{equation}
For ideal Poisson OOK, Eq.~\eqref{eq:detector_aware_map} reduces to the scalar
test in Eq.~\eqref{eq:ook_poisson_test}.  For coded reception, the useful output
is instead a log-likelihood ratio (LLR).  For bit $u_j$ in a modulation symbol,
\begin{equation}
 L_j(\mathbf y)=\ln
 \frac{\sum_{m:u_j(m)=0}p_{\mathcal D}(\mathbf y\mid m)P_X(m)}
      {\sum_{m:u_j(m)=1}p_{\mathcal D}(\mathbf y\mid m)P_X(m)} .
 \label{eq:photon_counting_llr}
\end{equation}
This form accommodates OOK, multilevel PAM, PPM, spatial diversity, and
multi-channel reception without first forcing the data into an additive-noise
model \cite{Gurugopinath2016Detection,Det0388,Det0077}.  When dead time or channel dispersion makes the
observation depend on previous symbols, sequence detection replaces the
memoryless rule.  Generalized ML sequence detection and multiple-symbol
detection improve robustness to unknown intensity and slowly varying channels,
although their state space grows with memory length and modulation order
\cite{Chatzidiamantis2010MLSD,Riediger2007MSD,Det0167}.

Model mismatch appears directly as LLR miscalibration.  A Poisson threshold
applied after SPAD blocking can become biased toward low-amplitude symbols, while
SiPM correlated avalanches make nominal Poisson LLRs overconfident.  Simple
variance-stabilizing or nonlinear equalizers can be effective in a restricted
operating region, and learned boundaries can track a multilevel transfer curve,
 but training data must cover background, temperature, and count-rate drift
 \cite{Arya2022MLBoundaries,Det0016,Det0158}.  SiPM equalization and neural
 detectors provide additional calibration examples \cite{Zhang2023SiPMEqualization,Det0025}.  The learned or equalized
 output should consequently be calibrated against held-out count data before it
 is passed to a soft decoder.
 Robust implicit-CSI detection, learned underwater equalization, and MIMO
 photon-counting receiver design provide three representative paths from a
 nominal likelihood to an implementable detector
 \cite{Lit0132,Lit0182,Lit0186}.

\subsubsection{Joint channel-gain and background estimation}

A balanced two-level pilot makes signal and background identifiable.  For the
$J$-pixel SPAD array used below, the off- and on-pilot gate counts obey
\begin{equation}
 C_i\sim\operatorname{Binomial}(J,p_i),\qquad
 p_i=1-\exp\!\left[-\frac{b+\eta g x_i}{J}\right].
 \label{eq:spad_pilot_occupancy}
\end{equation}
If $\bar C_0$ and $\bar C_1$ are the off- and on-pilot sample means, a
background-subtracted Poisson estimator is
\begin{equation}
 \widehat g_{\rm P}=\left[
 \frac{\bar C_1-\bar C_0}{\eta x_1}\right]^+,
 \label{eq:poisson_gain_estimator}
\end{equation}
whereas inversion of the finite-pixel response gives
\begin{equation}
 \widehat g_{\rm occ}=\left[
 \frac{-J\ln(1-\bar C_1/J)+J\ln(1-\bar C_0/J)}
 {\eta x_1}\right]^+ .
 \label{eq:occupancy_gain_estimator}
\end{equation}
The latter is the joint maximum-likelihood solution for the balanced,
independent-gate model.  Numerical optimization of
Eq.~\eqref{eq:joint_receiver_likelihood} generalizes it to unequal pilots,
nonuniform pixels, fading, and detector memory.  Channel estimation for
photon-counting and PMT receivers, operation under emulated fading, and adaptive
background rejection demonstrate why both $g$ and $b$ must be tracked
\cite{Det0138,Det0392,Det0134,Pachedjieva2018SpaceAccuracy}.

The common Monte Carlo study uses a 128-symbol balanced OOK pilot and
$5\times10^4$ trials per energy point.  The ideal, SNSPD, CMOS-SPAD, SiPM, and
MCP-PMT models use PDEs $(1,.8,.35,.30,.20)$ and resolving-cell counts
$(\infty,16,256,3600,64)$; the normalized values illustrate reported receiver
regimes rather than rank products \cite{Zhang2018APDSPADComparison,He2022SiPMSelection,Hao2024SNSPDArray,Guardiani2024SNSPDLaserCom}.

\begin{figure}[!t]
  \centering
  \includegraphics[width=\columnwidth]{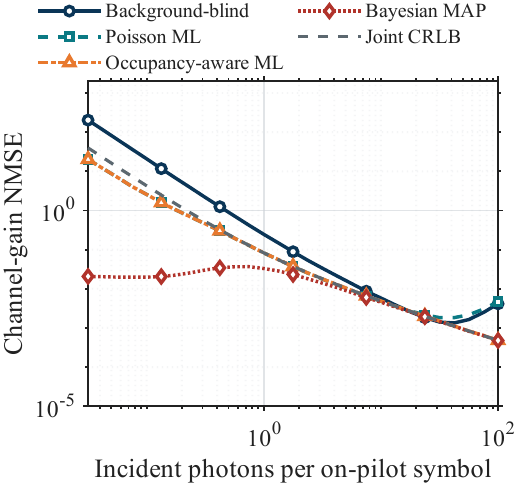}
  \caption{Channel-gain NMSE for four estimators and the joint CRLB using the
  same 256-pixel CMOS SPAD array.  Ordinary Poisson ML becomes biased by
  finite-pixel occupancy, whereas the detector-aware likelihood remains useful
  into the nonlinear regime.}
  \label{fig:channel_estimator_comparison}
\end{figure}

Figure~\ref{fig:channel_estimator_comparison} first holds the detector fixed and
changes only the estimator.  The ordinate is
$\operatorname{NMSE}=\mathbb E[(\widehat g-g)^2]/g^2$.  The background-blind
estimate is strongly biased at low pilot energy.  Background-subtracted Poisson
ML improves rapidly, but bends upward at high flux because it interprets finite
pixel occupancy as a linear count.  Occupancy-aware ML follows the joint
Cram\'er--Rao lower bound (CRLB) and, at the highest simulated energy, reduces
NMSE by a factor of 9.75 relative to Poisson ML.  The Bayesian MAP estimate uses
a truncated Gaussian prior; it reduces
low-count variance but retains a prior-dependent bias, and should not be
interpreted as universally superior.  Each marker is obtained from
$5\times10^4$ independent pilot blocks.  Aggregate binomial draws use a Poisson
approximation only when the individual firing probability is below 0.02 and the
mean is below 40; otherwise a bounded moment-matched Gaussian draw avoids
expanding hundreds of thousands of Bernoulli variables.

\subsubsection{One estimation framework across detector technologies}

The reverse comparison holds the algorithmic workflow fixed and changes the
front end.  For a detector with $R$ independent resolving cells per gate, the
mean primary count is represented by
\begin{equation}
 m_{\mathcal D}(\mu)=
 \begin{cases}
   \mu, & R=\infty,\\
   R\left(1-e^{-\mu/R}\right), & R<\infty,
 \end{cases}
 \label{eq:detector_mean_response_estimation}
\end{equation}
with $\mu=\eta(b_{\rm in}+gx)+d$.  The same joint-ML workflow estimates the
off- and on-pilot response and applies the inverse of
Eq.~\eqref{eq:detector_mean_response_estimation}.  For the SiPM case, the mean
primary response is additionally multiplied by 1.14 and the simulation draws
correlated events conditionally on the primary occupancy.  This simplified
branching term captures the estimator penalty from excess count variance; a
product-specific design would replace it with measured optical-crosstalk,
afterpulse, and recharge kernels.  SiPM mode selection and dual-readout studies
show that the correct observation model can change with the same physical
device \cite{Det0398,Li2025SiPMModes}.

Under common parameters, the ideal Poisson model is an algorithmic reference,
not a realizable detector.  The SNSPD benefits from high PDE and low dark count
at low and moderate pilot energies, but its illustrative 16 resolving modes per
gate eventually fill, producing a high-flux NMSE upturn.  The 256-pixel SPAD
array and 3600-cell SiPM tolerate greater instantaneous occupancy; the SiPM is
penalized by dark and correlated counts, while the representative MCP-PMT is
primarily PDE limited.  Parallel SPAD and SNSPD receivers can move these
boundaries by increasing the number of simultaneously available elements
\cite{Det0164,Gramuglia2021LowNoiseSPAD,Liu2025NovelCMOSSPAD,Chen2024SPADReceiver}.  This is a controlled comparison of response physics,
not a ranking of all commercial devices.

\subsubsection{Synchronization and joint acquisition}

For an oversampled preamble $a_k(\tau)$, known-parameter ML timing evaluates
\begin{equation}
 \widehat\tau_{\rm ML}=\arg\max_{\tau\in\mathcal G}
 \sum_k\ln p_{\mathcal D}
 \bigl(y_k\mid b+g a_k(\tau)\bigr),
 \label{eq:timing_ml}
\end{equation}
over a timing grid $\mathcal G$.  Maximum-bin acquisition discards the preamble
pattern; count correlation uses the pattern but not the heteroscedastic
likelihood.  When $g$ and $b$ are unknown, the GLRT replaces them for every
candidate $\tau$ by their conditional estimates.  The added nuisance parameters
produce a finite-sample acquisition penalty, but the joint metric avoids the
larger failure caused by a fixed, badly mismatched threshold.  Clock-recovery,
 time-coordinated SPAD reception, PPM symbol synchronization, and photon-counting
 time transfer all exploit related event-timing structure
 \cite{Det0014,Det0006,Det0026}.  Learned and time-gated receivers provide
 complementary event-timing implementations \cite{Det0156,Huang2021TimeGated}.
 Underwater PPM acquisition, sandwich-sequence design, likelihood-based PPM
 timing, and early photon-counting synchronizers provide complementary
 implementations \cite{Lit0008,Lit0022,Lit0160,Lit0170}.

A controlled comparison uses four algorithms on identical realizations from
the 256-pixel SPAD model.  The pulse occupies
three samples with weights $(0.25,1,0.40)$, so Poisson/binomial likelihood
weighting is not exactly equivalent to an unweighted correlation.  At each of
17 SBR values, $2\times10^4$ complete frames are generated.  The aggregate
Poisson approximation is valid here because the largest individual-pixel firing
probability is below 0.004.  All methods approach the random-acquisition RMSE at
low SBR.  In the waterfall region, exploiting the full preamble provides a clear
gain over maximum-bin selection; known-parameter ML is best, and the joint GLRT
approaches it while estimating gain and background from the same preamble.  The
small remaining difference between count correlation and ML is expected in the
 low-occupancy regime and grows for nonuniform background, broader pulse shapes,
 or saturated pixels.  Spatial tracking and centroiding introduce the same
 tradeoff in two dimensions \cite{Bashir2021BeamTracking,Srinivasan2018Centroiding,Licht2025PhotonCountingCamera}.
 Photon-array beam estimation, learning-based position recovery, Kalman tracking,
 and U-OFDM timing illustrate how the estimator changes with the measurement
 representation \cite{Lit0088,Lit0106,Lit0294,Lit0248}.

The practical receiver should therefore use a hierarchy.  A low-complexity
sliding statistic first identifies candidate frames; detector-aware joint
estimation then refines timing, gain, and background; finally, the calibrated
likelihood supplies soft data decisions.  Updates should be repeated on the
timescale of scintillation, pointing drift, ambient-light change, and detector
temperature, but not so frequently that pilot overhead dominates.  Receiver
experiments using automatic attenuation, dynamic detection cycles, clock/data
 recovery, and nonlinear compensation illustrate different implementations of
 this hierarchy \cite{Det0091,Det0123}.  Robust detection, improved underwater
 synchronization, self-synchronized UV reception, and joint communication--ranging
 acquisition extend the same hierarchy \cite{Lit0153,Lit0159,Lit0223,Lit0304}.
 The central design rule is that DSP can compensate
 for a known and identifiable response; it cannot reconstruct distinctions that
 were erased by saturation, unresolved simultaneous events, or an inadequately
 sampled timing waveform.
 Industrial VLC, high-rate micro-LED links, NOMA parameter estimation, and
 system-level synchronization studies provide useful stress cases for this
 principle \cite{Lit0178,Lit0221,Lit0275,Lit0255}.

\subsection{Channel Coding and Advanced Receiver Processing}
\label{sec:coding_receiver_processing}

Channel coding is most effective in a photon-counting receiver when it begins
before the threshold decision.  The count vector, detector state, channel
estimate, and synchronization confidence jointly determine the reliability of
each coded bit.  Converting these observations to one-bit decisions throws away
precisely the information that a soft decoder can exploit.  Recent experiments
 have therefore combined multichannel photon counting with modified LLRs and
 LDPC decoding, demonstrating that detector-aware processing is relevant from
 underwater links to high-speed SNSPD receivers
 \cite{Zhang2024HighSpeedPhC,Lit0042,Lit0157}.  The same issue arises for
 finite-array receivers, where missed photons, false counts, and blocking
 events need not have equal probabilities \cite{Lit0113,Lit0381}.  This subsection first establishes the
soft-information interface, then uses a reproducible QC-LDPC benchmark to
separate algorithmic and detector effects, and finally maps the result to
practical coding architectures.

\subsubsection{Detector-aware soft information}

Using the LLR convention of Eq.~\eqref{eq:photon_counting_llr}, positive values
favor coded bit zero.  For equiprobable OOK observed through an ideal Poisson
counter with means $\lambda_0$ and $\lambda_1$, the decoder input is
\begin{equation}
 L_{\rm P}(y)=y\ln\!\left(\frac{\lambda_0}{\lambda_1}\right)
              +(\lambda_1-\lambda_0).
 \label{eq:poisson_decoder_llr}
\end{equation}
For a gated array with $J$ independently available pixels and firing
probabilities $p_0$ and $p_1$, the matched metric is instead
\begin{equation}
 L_{\rm occ}(y)=y\ln\!\left(\frac{p_0}{p_1}\right)
 +(J-y)\ln\!\left(\frac{1-p_0}{1-p_1}\right),
 \label{eq:occupancy_decoder_llr}
\end{equation}
where $p_i=1-\exp(-\lambda_i/J)$.  Equation
\eqref{eq:occupancy_decoder_llr} retains the one-click-per-pixel constraint and
converges to Eq.~\eqref{eq:poisson_decoder_llr} at low occupancy.  A Gaussian
count approximation can be useful for large arrays, but its mean and variance
must remain symbol dependent; imposing equal variance erases the
input-dependent shot-noise structure.  Correlated SiPM avalanches, detector
recovery, and array nonuniformity require either a richer analytic law or an
empirical probability mass function calibrated from pilots
\cite{Sarbazi2018SPADStatistics,Zhang2023SiPMEqualization,Det0016,Det0025}.  Reliability calibration is consequently part of the receiver, not
an optional decoder tuning step.

When parameters are estimated, a plug-in LLR
$L(y;\widehat g,\widehat b,\widehat{\boldsymbol\theta}_{\mathcal D})$ is the
simplest interface to the FEC decoder.  Its magnitude should be clipped to avoid
numerical domination by a small number of badly modeled samples.  A more robust
alternative averages the likelihood over the posterior uncertainty of the
nuisance parameters,
\begin{equation}
 L_{\rm B}(y)=\ln
 \frac{\int p_{\mathcal D}(y\mid 0,\boldsymbol\xi)
 p(\boldsymbol\xi\mid\mathbf y_{\rm p})\,d\boldsymbol\xi}
 {\int p_{\mathcal D}(y\mid 1,\boldsymbol\xi)
 p(\boldsymbol\xi\mid\mathbf y_{\rm p})\,d\boldsymbol\xi},
 \label{eq:bayesian_detector_llr}
\end{equation}
where $\boldsymbol\xi$ collects gain, background, timing, and detector
parameters and $\mathbf y_{\rm p}$ denotes pilot observations.  Exact
marginalization is rarely needed in hardware: a short quadrature rule,
 confidence-dependent LLR scaling, or a small lookup table can capture most of
 the benefit.  Iterative multiuser processing under input-dependent shot noise
 extends the same principle by repeatedly updating interference and posterior
 symbol probabilities \cite{Lit0017,Lit0030,Lit0188}.

\subsubsection{QC-LDPC benchmark and simulation methodology}

 The first coded experiment holds the detector fixed and changes only the
 information supplied to the decoder.  A rate-$1/2$ quasi-cyclic LDPC code with
 $(N,K)=(1944,972)$ and circulant size $Z=81$ is used as a transparent,
 hardware-oriented benchmark.  The parity-check structure is used only as a
 reproducible algorithmic reference and is not presented as an optical-link
 standard.  Decoding uses layered belief
propagation, at most 30 iterations, early parity-check termination, and LLR
clipping at $\lvert L\rvert=30$.

For a fair energy comparison, the abscissa is incident photons per
\emph{information} bit, $E_{\rm ph,b}$.  With equiprobable OOK and code rate
$R_c$, the coded on-symbol energy is
\begin{equation}
 n_{\rm on}=2R_c E_{\rm ph,b},
 \label{eq:coded_ook_energy_normalization}
\end{equation}
whereas uncoded OOK uses $n_{\rm on}=2E_{\rm ph,b}$.  Thus redundancy is paid
for in the optical-energy accounting and is not hidden by plotting photons per
transmitted coded bit.

\begin{figure}[!t]
  \centering
  \includegraphics[width=\columnwidth]{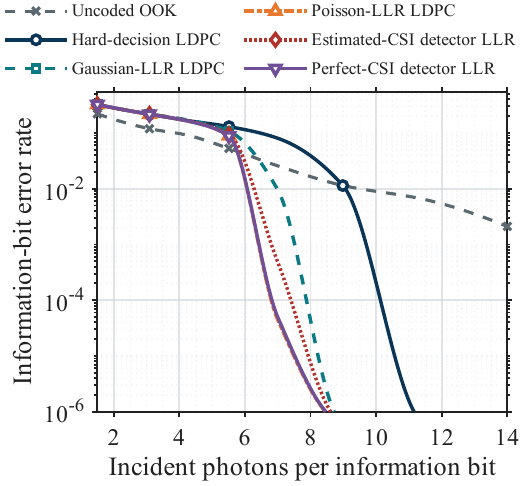}
  \caption{Information-bit BER for uncoded OOK and five rate-$1/2$ QC-LDPC
  receiver metrics using the same 32-cell SPAD model.  Markers are Monte Carlo
  results; lines are shape-preserving guides. Energy is normalized per
  information bit.}
  \label{fig:ldpc_likelihood_comparison}
\end{figure}

Figure~\ref{fig:ldpc_likelihood_comparison} compares six receiver outputs on
the same SPAD front end.  At low photon number, all coded methods are dominated
by the overlap between the off- and on-count distributions.  In the waterfall
region, hard-decision LDPC is clearly inferior because thresholding removes the
distance of each count from the decision boundary.  The Gaussian metric
recovers part of the loss, while the Poisson and finite-occupancy LLRs are
nearly coincident until array blocking becomes appreciable.  The
estimated-CSI occupancy metric approaches the perfect-CSI curve but retains a
visible pilot-estimation penalty near the waterfall.  The uncoded reference
decreases smoothly but lacks the steep coding transition.  These curves show
why the FEC gain cannot be specified independently of the detector likelihood.

The lowest displayed BER values are finite-simulation resolution points, not
claims of error-free operation.  At a high-confidence design target, the
waterfall simulation should be followed by importance sampling, trapping-set
analysis, or substantially longer hardware-in-the-loop runs.  This distinction
is especially important for optical transport links, where post-FEC targets can
be many orders of magnitude below the range accessible to direct Monte Carlo.
 Concatenated soft-inner/hard-outer architectures illustrate one route to low
 error floors with bounded memory and latency, while nonbinary and irregular
 LDPC designs provide alternative complexity--gain tradeoffs
 \cite{Lit0358,Lit0371,Lit0373}.

\subsubsection{One soft decoder across detector technologies}

The complementary experiment fixes the code, layered decoder, and matched-LLR
workflow, then changes only the detector response.  This isolates the front-end
penalty under the representative parameters above.  The ideal Poisson counter is a
reference bound.  The SNSPD model follows it most closely because of high PDE
and low dark count.  The CMOS SPAD shifts the waterfall to higher incident
energy primarily through its lower PDE, while its large pixel count keeps
blocking modest over the plotted range.  The representative SiPM combines low
PDE, dark events, and correlated avalanches; the MCP-PMT is mainly limited by
its lower effective detection efficiency.  These are controlled model
comparisons, not universal rankings: parallelization, wavelength, fill factor,
hold-off, optical coupling, and readout architecture can all reorder actual
devices \cite{Hao2024SNSPDArray,Guardiani2024SNSPDLaserCom,Chen2024SPADReceiver,He2022SiPMSelection}.

The comparison also exposes a useful design rule.  Coding can average random
count fluctuations, but it cannot restore distinctions erased by saturation.
If both OOK levels produce nearly the same occupied-cell count, the matched LLR
itself approaches zero and no decoder can recover the missing information.
Additional parallel pixels, shorter gates, adaptive attenuation, or a lower
code-symbol rate must first move the detector back into an identifiable region
\cite{Det0385,Li2022IntegratedSPAD,Liu2025NovelCMOSSPAD}.  Conversely, in the
low-occupancy region, a Poisson LLR may be sufficient and avoids needless model
complexity.

\subsubsection{Selecting the coding architecture}

 No single FEC family simultaneously maximizes photon efficiency, spectral
 efficiency, throughput, and minimum latency.  In a severely photon-starved
 downlink, high-order PPM concentrates energy into one slot and serially
 concatenated PPM (SCPPM) adds an outer convolutional code, interleaving, an
 accumulator, and iterative demapping/decoding.  PPM receiver studies spanning
 deep-space algorithms, coded multipulse modulation, and photon-information
 efficiency show that the interleaver, timing loop, detector interface, and
 decoder must be sized together \cite{Lit0172,Lit0195,Lit0214}.

 At higher photon flux, OOK avoids the PPM bandwidth expansion and pairs
 naturally with QC-LDPC codes.  Structured parity-check matrices and scheduled
 message passing can support highly parallel implementations.  For links
 whose dominant constraint is decoder latency or an ultra-low error floor, a
 short outer BCH/RS/product code can be concatenated with a soft inner code.  Polar and
 nonbinary LDPC concatenations are additional candidates at short-to-moderate
 block lengths, but they should be evaluated with the actual count likelihood
 rather than an AWGN surrogate \cite{Lit0367,Lit0358}.

\subsubsection{Advanced iterative and adaptive processing}

The natural extension is a turbo-style receiver in which synchronization,
channel estimation, demapping, and decoding exchange reliability information.
Decoder extrinsic probabilities can refine the gain and background estimate;
the updated likelihood then produces better LLRs for the next decoding pass.
For PPM, the same loop can include slot-timing and symbol constraints.  For
 dispersive underwater or indoor channels, a count-domain equalizer or sequence
 detector can exchange soft information with the FEC decoder.  Multi-channel
 photon-counting experiments already demonstrate the value of correcting LLRs
 for detector randomness \cite{Zhang2024HighSpeedPhC}, while asynchronous and
 PPM receivers demonstrate FPGA-oriented acquisition and likelihood-based slot
 synchronization \cite{Lit0099,Lit0160}.

Adaptation should be driven by quantities the receiver can measure: estimated
background, occupied-cell fraction, decoder syndrome trajectory, pilot
residuals, and timing confidence.  A practical controller can switch among
precomputed LLR tables, code rates, PPM orders, repetition factors, and hold-off
settings.  Learned equalizers are useful when analytic detector memory is too
complex, but they should output calibrated probabilities and retain a safe
fallback model outside the training range
\cite{Arya2022MLBoundaries,Det0158,Zhang2023SiPMEqualization}.  The final
engineering objective is therefore not the lowest uncoded BER or the strongest
stand-alone code.  It is the lowest required incident photons per delivered
information bit subject to bandwidth, latency, decoder power, acquisition
time, and a detector operating point that remains identifiable.

\section{System-Level Co-Design, Experimental Demonstrations, and Future Directions}
\label{sec:system_level_codesign}

\subsection{Photon-Counting Link Budget and System-Level Co-Design}
\label{sec:link_budget_codesign}

\begin{figure*}[!t]
  \centering
  \includegraphics[width=\textwidth]{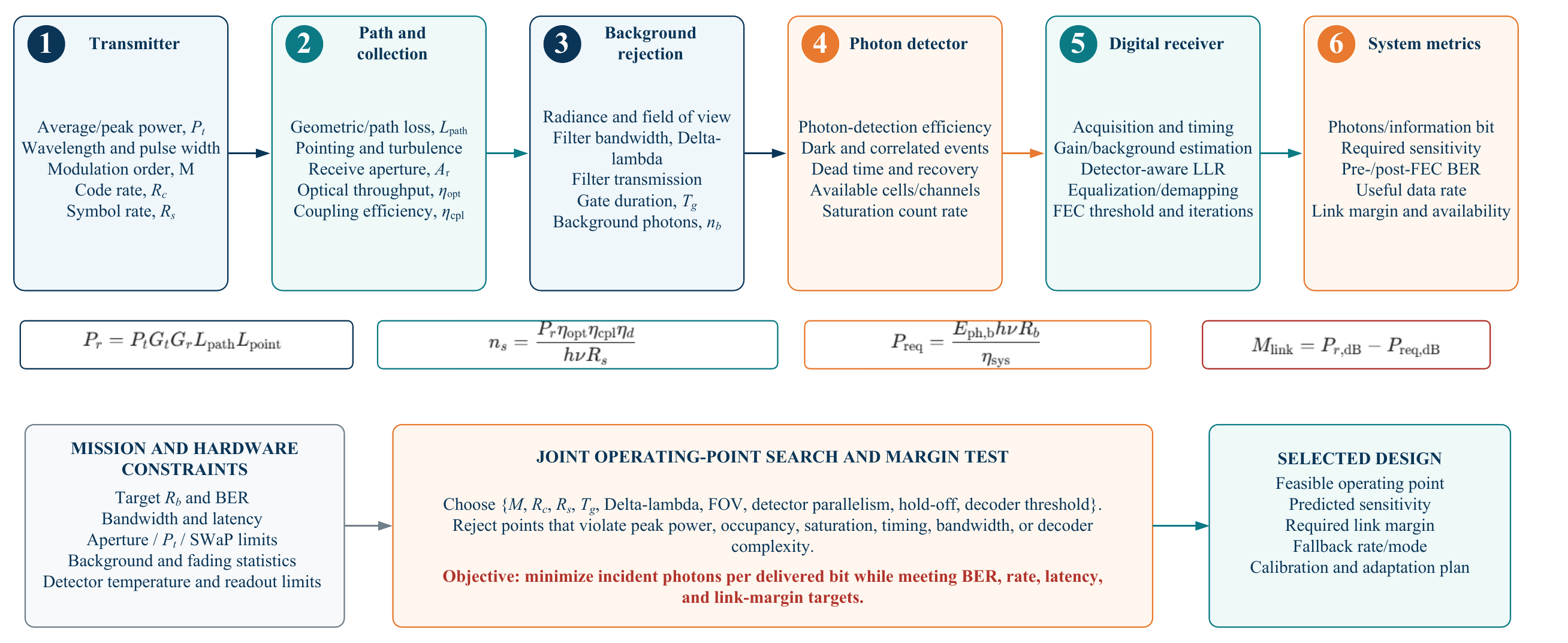}
  \caption{End-to-end photon-counting link-budget and system co-design flow.
  The upper path converts transmitted optical power into collected photons,
  registered events, decoded information, and system metrics.  The lower path
  rejects operating points that violate rate, background, detector occupancy,
  saturation, timing, bandwidth, latency, or post-FEC BER constraints, and
  feeds the required changes back to the optical and digital subsystems.}
  \label{fig:link_budget_codesign_flowchart}
\end{figure*}

A conventional optical link budget stops at received power.  A
photon-counting link budget must continue through optical filtering, coupling,
photon-detection efficiency, event loss, synchronization, and the post-FEC
decision threshold.  The relevant question is not merely whether photons reach
the aperture, but whether a sufficient number of statistically distinguishable
events reach the decoder without driving the detector into blocking or
saturation.  Figure~\ref{fig:link_budget_codesign_flowchart} organizes this
end-to-end calculation and closes the loop from decoded BER and link margin
back to transmitter power, modulation, code rate, gate width, filter bandwidth,
detector parallelism, and receiver processing.  This system view is consistent
with recent photon-counting ground-receiver development, emulated fading tests,
and high-dynamic-range receiver architectures
\cite{Downey2024GroundReceiver,Marsden2024GroundReceiver,Rogalin2021HighDynamicRange,Det0392}.

\subsubsection{Transmitted power, propagation, and collected signal}

Let $P_{\rm t}$ denote average launched optical power.  A convenient
factorization of the instantaneous power delivered to the detector-side optical
interface is
\begin{equation}
 P_{\rm r}=P_{\rm t}\eta_{\rm tx}L_{\rm prop}\eta_{\rm point}
 \eta_{\rm ap}\eta_{\rm cpl}\eta_{\rm rx},
 \label{eq:end_to_end_received_power}
\end{equation}
where $\eta_{\rm tx}$ and $\eta_{\rm rx}$ collect transmitter and receiver
optical throughput, $L_{\rm prop}$ includes deterministic propagation loss,
$\eta_{\rm point}$ represents pointing loss, $\eta_{\rm ap}$ is the fraction
intercepted by the receive aperture, and $\eta_{\rm cpl}$ is the coupling
efficiency into the detector or fiber interface.  All factors in
Eq.~\eqref{eq:end_to_end_received_power} are linear and no larger than unity;
their decibel losses add.  Atmospheric fading and pointing jitter should be
represented by distributions rather than hidden inside one average factor when
link availability is required.

For a centered Gaussian spot with $1/e^2$ radius $w_{\rm r}$ and circular
aperture radius $a_{\rm r}$,
\begin{equation}
 \eta_{\rm ap}=1-\exp\!\left(-\frac{2a_{\rm r}^{2}}{w_{\rm r}^{2}}\right).
 \label{eq:gaussian_aperture_collection}
\end{equation}
Equation~\eqref{eq:gaussian_aperture_collection} makes the aperture tradeoff
explicit: increasing $a_{\rm r}$ improves signal collection, but often also
increases background etendue, mass, pointing sensitivity, and coupling
complexity.  Single-mode coupling additionally depends on wavefront quality,
whereas multimode fibers and photonic lanterns exchange modal acceptance for
background and detector-channel count.  Ground-receiver and photonic-lantern
studies illustrate that aperture gain cannot be separated from coupling and
tracking performance \cite{Srinivasan2023DSOCReceiver,Tedder2020PhotonicLantern,Bashir2021BeamTracking}.

The mean number of detected signal-generated primary events in a gate of
duration $T_{\rm g}$ is
\begin{equation}
 \lambda_{\rm s}=\eta_{\rm d}\frac{P_{\rm r}T_{\rm g}}{h\nu},
 \label{eq:registered_signal_photons_gate}
\end{equation}
where $\eta_{\rm d}$ is PDE at optical frequency $\nu$.  This definition places
all receiver optical losses before $P_{\rm r}$ and only the photon-to-event
conversion in $\eta_{\rm d}$, preventing coupling loss from being counted
twice.  For pulsed PPM or return-to-zero signaling, $P_{\rm r}T_{\rm g}$ is
replaced by the received pulse energy admitted by the timing gate.

\subsubsection{Background radiance, filtering, and gating}

For approximately uniform spectral radiance $L_{\lambda}$ across the passband,
the background power admitted by a direct-detection receiver can be written
\begin{equation}
 P_{\rm b}\simeq L_{\lambda}A_{\rm r}\Omega_{\rm FOV}
 \Delta\lambda_{\rm eff}\eta_{\rm opt,b},
 \label{eq:background_radiance_power}
\end{equation}
where $A_{\rm r}$ is receive-aperture area, $\Omega_{\rm FOV}$ is accepted solid
angle, $\Delta\lambda_{\rm eff}$ is the noise-equivalent optical-filter
bandwidth, and $\eta_{\rm opt,b}$ is the background-path throughput.  The
registered background and dark-event mean per gate is then
\begin{equation}
 \lambda_{\rm b}=\eta_{\rm d}\frac{P_{\rm b}T_{\rm g}}{h\nu}
 +r_{\rm dc}T_{\rm g}.
 \label{eq:registered_background_gate}
\end{equation}
Equations~\eqref{eq:background_radiance_power} and
\eqref{eq:registered_background_gate} expose four powerful rejection levers:
spectral bandwidth, field of view, spatial-mode acceptance, and gate duration.
Narrowband filtering is particularly important in daylight photon-counting
FSO links \cite{Shtyrkova2019Narrowband}; however, a narrower filter can add
insertion loss, temperature sensitivity, angle dependence, or pulse distortion.
Likewise, a short gate reduces background linearly only if synchronization
jitter and pulse broadening do not exclude a significant fraction of signal
energy.  The correct optimization therefore maximizes a post-detection metric,
not filter rejection alone.

The signal-to-background ratio $\lambda_{\rm s}/\lambda_{\rm b}$ is not a
complete receiver metric.  Two links with equal ratio but different absolute
counts can have different dead-time loss, acquisition time, and decoder
reliability.  Background also changes the available dynamic range by occupying
pixels or recovery opportunities before signal photons arrive.  These effects
are visible in practical receiver analyses and background-illuminated SiPM
measurements \cite{Det0092,Det0142,Sarbazi2018SPADStatistics}.

\subsubsection{PDE, dead time, saturation, and decoding threshold}

The primary-event mean $\lambda=\lambda_{\rm s}+\lambda_{\rm b}$ must be
converted through the nonideal detector model before BER is evaluated.  For
$J$ independent one-click cells in a gate, the mean registered occupancy is
\begin{equation}
 \bar y_{\rm occ}=J\left[1-\exp\!\left(-\frac{\lambda}{J}\right)\right].
 \label{eq:system_occupancy_budget}
\end{equation}
For a nonparalyzable detector with $N_{\parallel}$ statistically independent
channels and per-channel dead time $\tau_{\rm d}$, a useful continuous-rate
approximation is
\begin{equation}
 r_{\rm out}\simeq
 \frac{r_{\rm in}}{1+r_{\rm in}\tau_{\rm d}/N_{\parallel}}.
 \label{eq:parallel_deadtime_rate_budget}
\end{equation}
Equations~\eqref{eq:system_occupancy_budget} and
\eqref{eq:parallel_deadtime_rate_budget} describe related but not identical
limits: the first is a gated occupancy constraint, while the second describes
recovery-limited throughput.  Afterpulsing, optical crosstalk, readout
arbitration, and pulse pileup further modify the count distribution.  A safe
design normally maintains headroom below both the occupancy knee and the
aggregate saturation count rate instead of relying on a decoder to repair
 information already compressed by the detector.  Parallel SNSPD and SPAD-array
 receivers demonstrate how $N_{\parallel}$ can raise the usable count rate
 \cite{Hao2024SNSPDArray,Lit0024,Chen2024SPADReceiver}.

The decoding threshold must be computed using the final event distribution.
Define the required incident photon energy per information bit as
\begin{equation}
 E_{\rm ph,b}^{\rm req}=\inf\left\{E:
 P_{\rm b,post}(E,\lambda_{\rm b},\boldsymbol\theta_{\mathcal D},
 \mathcal M,\mathcal C)\leq P_{\rm target}\right\},
 \label{eq:required_photons_per_bit_definition}
\end{equation}
where $\boldsymbol\theta_{\mathcal D}$ denotes detector parameters,
$\mathcal M$ the modulation and symbol mapping, and $\mathcal C$ the code and
decoder.  This definition includes synchronization, channel-estimation, LLR
mismatch, finite block length, and implementation loss.  It also prevents an
uncoded sensitivity number from being combined with an unrelated coding gain.

\subsubsection{Unified photons-per-bit, sensitivity, rate, and margin}

For an $M$-ary symbol stream at symbol rate $R_{\rm s}$ and code rate $R_{\rm
c}$, the delivered information rate is
\begin{equation}
 R_{\rm b}=R_{\rm s}R_{\rm c}\log_2M,
 \label{eq:useful_information_rate}
\end{equation}
provided that $R_{\rm s}$ is defined per complete modulation symbol.  For PPM,
the occupied bandwidth also scales with the number and width of slots, so the
same $R_{\rm b}$ does not imply the same bandwidth as OOK or PAM.  At the
detector optical interface, the incident photons per delivered information bit
are
\begin{equation}
 E_{\rm ph,b}=\frac{P_{\rm r}}{h\nu R_{\rm b}}.
 \label{eq:incident_photons_per_information_bit}
\end{equation}
If Eq.~\eqref{eq:required_photons_per_bit_definition} gives the required
incident value, the corresponding average sensitivity is
\begin{equation}
 P_{\rm req}=E_{\rm ph,b}^{\rm req}h\nu R_{\rm b},
 \qquad
 P_{\rm req,dBm}=10\log_{10}\!\left(\frac{P_{\rm req}}{1~{\rm mW}}\right).
 \label{eq:photon_sensitivity_conversion}
\end{equation}
Equations~\eqref{eq:parallel_deadtime_rate_budget} and
\eqref{eq:photon_sensitivity_conversion} show that low-rate sensitivity is
background limited, whereas high-rate feasibility is capped by aggregate count
rate.  Consequently, the best low-flux detector need not support the highest
useful information rate.

The definition point must always be stated.  ``Incident photons/bit'' before
PDE, ``detected photons/bit'' after PDE, and photons/bit inferred from an
electrical sensitivity measurement are not interchangeable.  Near-single-
photon-per-bit demonstrations and quantum-limit analyses provide important
benchmarks, but the usable system number must include optical loss, background,
detector response, and code overhead
\cite{Caplan2006PhotonEfficient,Dolinar2012UltimateLimits,Caplan2024PhotonCounting}.

For available received power $P_{\rm r,avail}$ at the selected outage
percentile, the link margin is
\begin{equation}
 M_{\rm link}=P_{\rm r,avail,dBm}-P_{\rm req,dBm}-L_{\rm impl,dB},
 \label{eq:photon_counting_link_margin}
\end{equation}
where $L_{\rm impl}$ includes losses not already represented in the physical or
Monte Carlo model.  A mean-power margin is insufficient for a turbulent or
blocked link; availability requires evaluating Eq.~\eqref{eq:photon_counting_link_margin}
over the received-power distribution and the adaptation timescale.

\subsubsection{Joint selection of detector, modulation, code, and rate}

The design variables can be collected as
$\mathbf x=(\mathcal D,\mathcal M,\mathcal C,R_{\rm c},R_{\rm s},T_{\rm g},
\Delta\lambda,\Omega_{\rm FOV},N_{\parallel})$.  A communication-oriented
selection problem is
\begin{align}
 \min_{\mathbf x}\quad &E_{\rm ph,b}^{\rm req}(\mathbf x) \\
 \text{subject to}\quad
 &P_{\rm b,post}\leq P_{\rm target},\quad R_{\rm b}\geq R_{\rm target},
 \nonumber\\
 &B(\mathbf x)\leq B_{\max},\quad
 r_{\rm out}(\mathbf x)\leq \rho r_{\rm sat},
 \nonumber\\
 &T_{\rm latency}(\mathbf x)\leq T_{\max},\quad
 P_{\rm elec}(\mathbf x)\leq P_{\max},
 \label{eq:system_codesign_optimization}
\end{align}
where $0<\rho<1$ reserves saturation headroom.  In practice the search is over
a finite set of detector channels, modulation orders, code rates, gate widths,
and filter options.  Each candidate first passes an optical collection and
background calculation, then a detector occupancy/dead-time test, and finally
a detector-aware demodulation and FEC simulation.  Only candidates satisfying
all three levels receive a link-margin score.

This procedure produces intuitive regime changes.  High-order SCPPM is favored
when photon efficiency dominates and bandwidth is available; soft-decision
OOK/QC-LDPC is attractive when bandwidth and throughput dominate; increasing
detector parallelism is preferable to stronger coding when saturation has
already erased amplitude information.  Wider filters or gates may be required
for large Doppler uncertainty and timing jitter, but their background penalty
 must be paid through aperture, optical power, coding, or reduced rate.  Published
 deep-space receiver algorithms, coded multipulse modulation, and optical LDPC
 studies similarly couple modulation, coding, synchronization, and hardware
 throughput rather than optimizing them independently
 \cite{Lit0172,Lit0214,Lit0358}.  The next subsection will compare reported experimental
systems using these common quantities.

\subsection{Experimental Demonstrations and Cross-System Benchmarking}
\label{sec:experimental_benchmarking}

Photon-counting links are often compared by quoting one record number, such as
maximum distance, peak data rate, sensitivity, or photons per bit.  Such a
comparison is incomplete because the reported number depends on the optical
reference plane, detector operating point, background, waveform overhead,
error criterion, and whether the propagation channel was physically traversed
or only emulated.  Table~\ref{tab:photon_counting_system_benchmarks} therefore
places representative space, terrestrial, and underwater systems on a common
reporting grid.  It includes both recent high-rate demonstrations and several
field or photon-efficiency landmarks needed to expose the underlying Pareto
front rather than to create a single league table.

Three rules were used when transcribing the results.  First, a physical field
link, a laboratory free-space path, an attenuation-equivalent water path, and
a circuit simulation are labeled separately.  Second, incident and detected
photons per bit are retained exactly as defined by the source.  Third, the
reported rate and error metric are not silently converted between uncoded,
pre-FEC, post-FEC, line, and delivered information quantities.  A dash therefore
means that a sufficiently comparable number was not reported, not that the
quantity was zero.

\begin{table*}[p]
\centering
\caption{Representative photon-counting optical communication demonstrations and receiver benchmarks.}
\label{tab:photon_counting_system_benchmarks}
\vspace{-0.5mm}
\fontsize{7.0}{7.9}\selectfont
\setlength{\tabcolsep}{1.80pt}
\renewcommand{\arraystretch}{1.02}
\begin{tabularx}{\textwidth}{@{}L{1.75cm}L{2.10cm}L{1.95cm}L{2.55cm}L{1.15cm}L{2.55cm}Y@{}}
\toprule
\textbf{Scenario and source} & \textbf{Evidence and channel} &
\textbf{Detector and wavelength} & \textbf{Waveform, coding, and processing} &
\textbf{Reported rate} & \textbf{Sensitivity or photon efficiency} &
\textbf{Background, error point, and interpretation} \\
\midrule
\multicolumn{7}{@{}l}{\textit{Space and long-haul field or emulated links}} \\
\addlinespace[0.3pt]

LLCD lunar downlink \cite{Grein2015LLCDReceiver} &
Field link; $\sim$230,000-km lunar path &
Four-element NbN SNSPD; 1550 nm & 16-PPM with SCPPM FEC; real-time reception & Up to 622 Mb/s &
$\sim$1.5 incident photons/bit at lower-rate modes &
Operational space link; required photons/bit rise at the highest rates because of detector blocking. \\

DSOC deep-space downlink \cite{Biswas2024DeepSpacePhoton,Srinivasan2023DSOCReceiver} &
Field link; reported operation through 0.56 AU &
64-wire WSi SNSPD ground receiver; 1550 nm & Variable-rate SCPPM and parallel event processing & 0.055--267 Mb/s &
Detector efficiency approximately 45--70\%, depending on flux &
The defining result is operational range and rate agility; high-flux blocking requires detector-aware mode selection. \\

Daytime CubeSat downlink \cite{Wan2024DaytimeSNSPD} &
Field link; 550-km LEO-to-ground path &
Dual-mode SNSPD receiver; 1550 nm & NRZ-OOK with count-threshold detection & 50 Mb/s &
$-45$ dBm with negligible background; $-35$ dBm near 10 Mcount/s background &
BER below $10^{-3}$ in the equivalent-sensitivity test; field power was typically $-55$ to $-40$ dBm. \\

Ground-receiver stress test \cite{Marsden2024GroundReceiver} &
Link emulation with fading, background, Doppler, and flux transients &
Up to 16 SNSPD channels; 1550 nm & 32-PPM, rate-$1/3$ CCSDS HPE SCPPM on RFSoC/FPGA & 21--83 Mb/s &
More than 28 dB demonstrated dynamic range &
$\pm66$-ppm clock offset produced less than 0.2-dB penalty; high-flux limit was detection delay, not shot noise. \\

\addlinespace[0.5pt]
\multicolumn{7}{@{}l}{\textit{Laboratory free-space receiver frontiers}} \\
\addlinespace[0.3pt]

Compact PNR SNSPD array \cite{Hao2024SNSPDArray} & Laboratory optical channel &
Four quadrants, 24 serial subpixels; 91.6\% DDE at 1550 nm & 8-PPM and rate-$1/2$ SCPPM; PNR soft information &
120 Mb/s--1.5 Gb/s &
1.47, 1.76, 3.40, and 7.41 incident photons/bit at 0.48, 0.8, 1.2, and 1.5 Gb/s &
1.61-Gcount/s capability; 0.8 background photons/slot was tolerated in the 120-Mb/s mode. \\

Psyche photon-counting camera \cite{Lit0043} & Laboratory receiver development for a deep-space ground terminal &
1024-pixel photon-counting focal plane & Spatial photon counting for acquisition, background rejection, and tracking & Not reported &
Not reported &
Device-level case emphasizing focal-plane scaling and event readout rather than a complete link BER. \\

InGaAs SPAD/SCPPM \cite{Lai2025InGaAsSPAD} & Laboratory free-space system &
$8\times8$ InGaAs SPAD array; 1550 nm & 4-PPM with SCPPM coding & 10 Mb/s &
5 incident photons/bit &
Recent noncryogenic telecom-band result; performance reflects array dead time and the selected iterative decoder. \\

Commercial SPAD high-rate link \cite{Huang2022FiveGbpsSPAD} & Short laboratory free-space path &
14,410-cell commercial SiPM/SPAD array; 405 nm & OFDM, nonlinear equalization, clipping, and adaptive bit/energy loading & 5 Gb/s &
Approximately $-35$ dBm at BER $2\times10^{-3}$ &
The peak-rate point relies on multicarrier loading and offline DSP; it is not a photon-efficiency record. \\

NIR SiPM link \cite{Ali2023SiPM82Gbps} & Laboratory; 35-cm free-space path &
1-mm$^2$ SiPM; 850 nm & OOK with decision-feedback equalization & 8.2 Gb/s &
15.34 W/m$^2$ irradiance at BER $3.8\times10^{-3}$ &
Similar result in darkness and under 500-lux white-LED illumination with long-pass filtering; shortest path in the table. \\

\addlinespace[0.5pt]
\multicolumn{7}{@{}l}{\textit{Underwater photon-counting links}} \\
\addlinespace[0.3pt]

SiPM UWOC diversity \cite{Det0020} & Water-tank experiment; 55-m physical path &
SiPM diversity receiver; blue--green band & OOK with nonlinear decision-feedback equalizer & 2 Gb/s &
$-41.96$ dBm &
Diversity and nonlinear equalization jointly extend the bandwidth/range operating region; not reported in photons/bit. \\

Long-range SiPM UWOC \cite{Zhang2020SiPMUWOC} & Water-tank experiment; 11.6 AL without turbulence and 9.28 AL with turbulence &
SiPM receiver; 450 nm & OOK with equalization & 1 Gb/s &
Source-reported receiver-power curves &
Attenuation lengths are more transferable than nominal distance, but the scattering and turbulence model must still be stated. \\

Photon-efficient PMT UWOC \cite{Hu2018UWOC256PPM} & Tank plus equivalent loss; 35.88 AL and 120-m Jerlov-II equivalent &
PMT photon counter; 532 nm & 256-PPM and RS(255,127) coding & 13.7 Mb/s &
3.32 information bits/detected photon &
136.8-dB link loss and symbol error rate $6.31\times10^{-4}$; the 120-m value is attenuation-equivalent. \\

Single-SPAD UWOC \cite{Lyu2022SPADUWOC} & Water-tank experiment; 7-m physical path &
Single SPAD module; 450 nm & QPSK-OFDM or 16-QAM-OFDM & 21.2 or 42.4 kb/s &
$-84.5$ dBm and 68 photons/symbol for QPSK; $-77.8$ dBm and 390 photons/symbol for 16-QAM &
Shows the sensitivity/spectral-efficiency tradeoff without conflating extrapolated reach with the physical tank length. \\

Multichannel photon-counting UWOC \cite{Zhang2024HighSpeedPhC} & Tank plus emulation; 94.09-dB loss and 459-m Jerlov-IA equivalent &
Parallel photon-counting channels; 450 nm & OOK, LDPC coding, and multichannel combining & 10.13 Mb/s &
$-85$ dBm and 0.358 incident photons/bit &
Scintillation index 0.0329; the extreme distance is an equivalent-channel result, not a 459-m physical tank. \\

\addlinespace[0.5pt]
\multicolumn{7}{@{}l}{\textit{Integrated-receiver simulation target}} \\
\addlinespace[0.3pt]

Integrated CMOS SPAD receiver \cite{Chen2024SPADReceiver} & Post-layout/circuit-level receiver simulation &
$128\times128$ CMOS SPAD array; visible band & Column-wise decoding and spot tracking & 500 Mb/s &
$-41$ dBm at BER $3\times10^{-4}$ &
Useful integration target with 100-kHz tracking, but it must not be plotted as an experimental link measurement. \\

\bottomrule
\end{tabularx}
\vspace{0.5mm}
\parbox{\textwidth}{\fontsize{6.1}{6.8}\selectfont
\textit{Notes:} DDE, device detection efficiency; HPE, high photon efficiency; PNR, photon-number resolving; AL, attenuation length.  Rates, powers, and photon efficiencies are source-reported values and may refer to different error criteria or optical reference planes.  ``Equivalent'' denotes calibrated attenuation or channel emulation rather than a physical propagation distance.}
\end{table*}

\subsubsection{Three performance regimes rather than one winner}

The table separates three distinct regimes.  Space and deep-space systems use
large receive apertures, narrow spatial and spectral acceptance, high-order PPM,
and powerful SCPPM coding to minimize photons per delivered bit.  Their rates
are governed jointly by slot width, detector blocking, FPGA throughput, Doppler
uncertainty, and the available downlink flux.  The LLCD and DSOC results are
therefore stronger evidence of an end-to-end photon-efficient architecture than
laboratory sensitivity alone, whereas the daytime CubeSat result demonstrates
the additional cost of background rejection and operational tracking
\cite{Grein2015LLCDReceiver,Biswas2024DeepSpacePhoton,Wan2024DaytimeSNSPD}.

Short terrestrial links occupy a different corner of the design space.  Large
SPAD/SiPM cell counts, wide electrical bandwidth, nonlinear equalization, and
adaptive multicarrier loading enable multi-gigabit throughput, but at received
fluxes well above the few-photon-per-bit regime
\cite{Huang2022FiveGbpsSPAD,Ali2023SiPM82Gbps}.  These demonstrations establish
detector and DSP bandwidth, not deep-space link sensitivity.  Conversely, the
 recent PNR-SNSPD results show that parallelism can move the photon-efficient
 frontier toward gigabit rates, although cryogenics, optical coupling, event
 transport, and saturation remain system-level costs
 \cite{Hao2024SNSPDArray,Lit0024,Lit0043}.

Underwater systems trade throughput against attenuation length, scattering,
and turbulence.  High-order PPM with a PMT achieves high detected-photon
efficiency at modest rate, SiPM diversity and equalization support gigabit
operation over a physical tank, and multichannel photon counting plus LDPC
coding produces extreme equivalent loss tolerance
\cite{Hu2018UWOC256PPM,Det0020,Zhang2024HighSpeedPhC}.  These are complementary
results.  A distance extrapolated from Beer--Lambert loss cannot be treated as
equivalent to a physical path when beam spread, volume scattering, alignment,
and water inhomogeneity are absent from the emulator.

\subsubsection{How to normalize a published result}

A useful first normalization is the average incident energy per delivered bit,
computed from Eq.~\eqref{eq:incident_photons_per_information_bit} at a declared
optical reference plane.  If only sensitivity is reported, wavelength and net
information rate are required for the conversion.  If PDE is then used to infer
detected photons per bit, its value must correspond to the same bias, count
rate, wavelength, and coupling state as the communication measurement.  This is
particularly important for SNSPD and SPAD arrays whose effective efficiency
falls or becomes nonlinear as channels block or pixels recover.

The second normalization is resource-aware.  At minimum, a benchmark should
pair photon efficiency with occupied bandwidth or slot rate, net throughput,
electrical power, aperture, detector active area or pixel count, and latency.
For background-limited links, filter bandwidth, field of view, gate width, and
background count rate are indispensable.  For coded systems, code rate, block
length, decoder iterations, and whether BER is pre- or post-FEC must be given.
For underwater and atmospheric links, physical distance should be reported
separately from attenuation-equivalent distance, along with turbulence or
scintillation statistics.  These fields are sufficient to reconstruct most of
the link budget and reveal which subsystem created the apparent gain.

\subsubsection{Benchmarking implications for system co-design}

No detector is universally dominant.  SNSPD arrays presently define the
low-background, telecom-band sensitivity frontier and support operational
space links, but require cryogenic packaging and parallel high-rate readout.
CMOS SPAD arrays offer dense integration, timing, imaging, and tracking, yet
their fill factor, dark counts, and dead-time compression must be included in
the communication model.  SiPMs provide large area and aggregate count rate at
low cost, but optical crosstalk, recovery, saturation, and equalizer dependence
become prominent at high flux.  PMT and upconversion receivers remain valuable
where aperture, wavelength conversion, or very low flux outweigh size and
integration constraints \cite{Det0049,Hu2018UWOC256PPM}.

Accordingly, the defensible comparison is a Pareto surface over incident
photons per delivered bit, spectral efficiency, net rate, range or link loss,
background tolerance, latency, and implementation complexity.  Future
demonstrations should publish the raw count statistics and a complete receiver
reference-plane budget in addition to the best BER curve.  This would allow a
detector, modulation, code, and synchronizer to be re-evaluated under a common
nonideal model instead of being ranked by incompatible headline metrics.

\subsection{Open Challenges, Standardization, and Research Directions}
\label{sec:future_directions}

Table~\ref{tab:photon_counting_system_benchmarks} shows that progress must now
be measured by sustained delivered information rate and availability across
acquisition, background rejection, event transport, synchronization, and FEC.

\subsubsection{Scaling from detector pixels to usable event throughput}

More pixels add area, rate, and spatial information but also wiring, routing,
skew, calibration, collisions, and power.  Hierarchical arrays should expose
selectable event, histogram, and decision modes suited to OOK, PPM, and tracking
\cite{Lit0079,Licht2025PhotonCountingCamera,Det0072,Liu2025NovelCMOSSPAD,Li2025SiPMModes,Det0398}.  The useful benchmark is sustained
information rate at stated event loss and electrical or cryogenic power.

\subsubsection{Background, pointing, and nonstationary channels}

Operational background, coupling, and occupancy are nonstationary.  A receiver
state vector containing background, timing confidence, spot position, and
coherence time should control gates, attenuation, hold-off, rate, and pixel
selection; field and underwater studies support this joint adaptation
\cite{Wan2024DaytimeSNSPD,Det0134,Det0392,Det0108,Det0378,Sun2025AirWater}.

\subsubsection{Integration, energy, and thermal constraints}

The appropriate objective is delivered information per joule under thermal and
latency limits, including cooling, bias, TDCs, I/O, memory, and decoding.  Local
filtering and histogramming can reduce event traffic; real-time systems should
report joules per decoded bit, bandwidth, latency, and utilization with BER
\cite{Vyhnalek2019Realtime,Lit0099}.

\subsubsection{Standards and reproducible benchmarking}

Reusable PPM and coding baselines should be extended by detector-aware profiles
covering reference planes, background, dead time, Doppler, timing drift, fades,
and acquisition \cite{Lit0170,Lit0177,Lit0358}.  Machine-readable settings and
representative count traces are especially important for learned receivers
\cite{Arya2022MLBoundaries,Det0382}.

\subsubsection{A staged experimental roadmap}

A staged roadmap proceeds from component characterization to hardware-in-loop,
physical-path, and extended-operation tests.  Larger arrays, adaptive SiPMs,
blocking-aware codes, hybrid coherent/counting receivers, and bidirectional
underwater links should be judged by availability, reacquisition, calibration,
and energy per delivered bit \cite{Mikos2026CoherentCombination,Caplan2024PhotonCounting,Det0037,Det0064,Det0382}.

\section{Conclusion}
\label{sec:conclusion}

Single-photon OWC sensitivity is an end-to-end property of propagation,
collection, filtering, coupling, detection, readout, synchronization, and
decoding.  The port-to-bit framework therefore declares a receiver reference
plane, retains detector nonidealities, and reports decoded BER, photon and
spectral efficiency, throughput, availability, and link margin.  No detector is
universally optimal: SPAD arrays favor integration and spatial timing, SiPMs
favor area and simple aggregation, SNSPDs favor low noise and timing at a
cryogenic cost, and PMT or upconversion receivers remain useful in specialized
wavelength and collection regimes.  PDE alone is insufficient when recovery,
occupancy, correlated noise, nonuniformity, and readout loss dominate.

Waveforms, estimators, synchronizers, and decoders must likewise match the
observed count or time-tag process; added DSP cannot restore distinctions erased
by saturation or event loss.  Future progress should be judged by reproducible
end-to-end demonstrations with common reference planes, background and fade
profiles, detector-state reporting, confidence intervals, and count traces.
Scalable arrays, adaptive gating and attenuation, reconfigurable readout, and
joint communication and pointing should ultimately be measured by sustained,
reliable information delivery under realistic power, latency, and calibration
constraints.

\bibliographystyle{IEEEtran}
\bibliography{full_references}

\end{document}